\documentclass[preprint, review,3p,times, 10pt]{elsarticle}

\usepackage{cite}
\usepackage{longtable}
\usepackage{amsmath,amssymb,amsfonts}
\usepackage{pifont}
\usepackage{multirow}
\usepackage{float}
\usepackage{placeins}
\usepackage[export]{adjustbox}
\usepackage{graphicx, caption, subcaption}

\usepackage{algorithmic}
\usepackage{graphicx}
\usepackage{textcomp}
\usepackage{comment}
\usepackage{footnote}
\usepackage{comment}
\usepackage{threeparttable}
\usepackage{romannum}
\usepackage{longtable}
\usepackage{lscape}
\usepackage{soul}
\usepackage{array, makecell}

\usepackage[hyphens]{url}
\usepackage{hyperref}
\hypersetup{colorlinks=true,breaklinks=true}

\newcommand{\quotes}[1]{``#1''}

\journal{''}

\begin{document}

\begin{frontmatter}




\title{Robust COVID-19 Detection from Cough Sounds using Deep Neural Decision Tree and Forest: A Comprehensive Cross-Datasets Evaluation}

\author[1]{Rofiqul Islam}
\ead{rofiqcsecu101@gmail.com}
\author[1]{Nihad Karim Chowdhury \corref{correspondingauthor}}
\cortext[correspondingauthor]{Corresponding authors.}
\ead{nihad@cu.ac.bd}
\author[2]{Muhammad Ashad Kabir\corref{correspondingauthor}}
\ead{akabir@csu.edu.au}%

\affiliation[1]{organization={Department of Computer Science and Engineering, University of Chittagong}, city={Chattogram},postcode={4331},country={Bangladesh}}
\affiliation[2]{organization={School of Computing, Mathematics, and Engineering, Charles Sturt University}, city={Bathurst}, state={NSW}, postcode={2795}, country={Australia}}


\begin{abstract}

This research presents a robust approach to classifying COVID-19 cough sounds using cutting-edge machine learning techniques. Leveraging deep neural decision trees and deep neural decision forests, our methodology demonstrates consistent performance across diverse cough sound datasets. We begin with a comprehensive extraction of features to capture a wide range of audio features from individuals, whether COVID-19 positive or negative. To determine the most important features, we use recursive feature elimination along with cross-validation. Bayesian optimization fine-tunes hyper-parameters of deep neural decision tree and deep neural decision forest models. Additionally, we integrate the synthetic minority over-sampling technique during training to ensure a balanced representation of positive and negative data. Model performance refinement is achieved through threshold optimization, maximizing the ROC-AUC score. Our approach undergoes a comprehensive evaluation in five datasets: Cambridge (asymptomatic and symptomatic), Coswara, COUGHVID, Virufy, and the combined Virufy with the NoCoCoDa dataset. Consistently outperforming state-of-the-art methods, our proposed approach yields notable AUC scores of 0.97, 0.98, 0.92, 0.93, 0.99, and 0.99, alongside remarkable precision scores of 1, 1, 0.72, 0.93, 1, and 1 across the respective datasets.
Merging all datasets into a combined dataset, our method, using a deep neural decision forest classifier, achieves an accuracy of 0.97, AUC of 0.97, precision of 0.95, recall of 0.96, F1-score of 0.96, and specificity score of 0.97.
Also, our study includes a comprehensive cross-datasets analysis, revealing demographic and geographic differences in the cough sounds associated with COVID-19. These differences highlight the challenges in transferring learned features across diverse datasets and underscore the potential benefits of dataset integration, improving generalizability and enhancing COVID-19 detection from audio signals.
\end{abstract}

\begin{keyword}
Cough sound\sep COVID-19 \sep Cross-datasets \sep Classification \sep Detection \sep RFECV \sep DNDF \sep DNDT \sep audio signal.
\end{keyword}
\end{frontmatter}

\section{Introduction}
\label{sec:introduction}

COVID-19 is a highly contagious disease caused by the SARS-CoV-2 virus. It can cause serious illness, death, and economic disruption. The surge in COVID-19 cases burdened medical diagnostic laboratories and healthcare facilities, underscoring the limitations of conventional diagnostic methods, such as clinical examinations, CT scans, and PCR tests \citep{CTandPCR}. Although PCR is a highly accurate method for COVID-19 detection, its cost and time requirements render it inaccessible to a significant portion of the population \citep{xiao2020}. Furthermore, the utility of CT scan imaging for COVID-19 diagnosis is constrained by the potential overlap of symptoms with influenza, delayed detection of lung manifestations in the early stages of infection, and restrictions on its use, notably for infants and pregnant women~\citep{KHORRAMDELAZAD2021104554,covidmanagement}. 

Studies have indicated the potential of using the human voice as a primary diagnostic tool for conditions associated with voice production, including the detection of pathological voice \citep{7472923}, pertussis \citep{pramono2023coughbased}, asthma \citep{doi:10.3109/03091902.2012.758322}, and respiratory diseases \citep{Swarnkar_Abeyratne_Chang_Amrulloh_Setyati_Triasih_2013}. These instances underscore the promising capabilities of the human voice in diagnosing diseases, particularly those related to voice production. Researchers have explored acoustic analysis to extend COVID-19 detection beyond voice-related conditions. They have investigated various modalities, including cough sounds, breathing patterns, and speech or voice, encompassing vowels. The initial two modalities present indicators for the symptoms of COVID-19, namely persistent coughing and breathlessness. We opted for the first modality due to the substantial volume of available cough data.  Numerous universities around the world, such as the University of Cambridge in the UK \citep{10.1145/3394486.3412865}, the Massachusetts Institute of Technology (MIT) in the United States \citep{Chaudhari2020VirufyGA}, and the École Polytechnique Fédérale de Lausanne (EPFL) in Switzerland \citep{Orlandic2020TheCC}, are actively involved in researching the application of machine learning (ML) techniques for the diagnosis of COVID-19 using cough sound data.

To expedite research focused on detecting COVID-19 through cough sound analysis, we introduce an approach that leverages deep neural decision trees (DNDT) and deep neural decision forests (DNDF). Initially, audio samples are preprocessed to extract various audio characteristics from cough sounds of individuals with confirmed positive and negative results of the COVID-19 test. Subsequently, we use Recursive Feature Elimination with Cross-Validation (RFECV) in combination with the Extra-Trees classifier to identify the most critical features for our classifiers. Following this, Bayesian optimization (BO) is used to fine-tune the hyper-parameters of our proposed method. To enhance the classification performance of our models, we incorporate threshold moving (TM) techniques to ascertain the optimal threshold value. The dataset prominently displays an imbalance, notably with a limited representation of positive instances for COVID-19, potentially posing a detrimental effect on the performance of the ML classifier. In response to this imbalance, we have implemented the synthetic minority over-sampling technique (SMOTE) \citep{SMOTE} during the training process. This strategic inclusion aims to rectify the dataset imbalance, thereby enhancing the performance of the ML classifier.

The evaluation of our classification models extends across multiple datasets, including Cambridge \citep{10.1145/3394486.3412865}, Virufy \citep{Chaudhari2020VirufyGA}, COUGHVID \citep{Orlandic2020TheCC}, Coswara \citep{Sharma2020CoswaraA}, and Virufy merged with NoCoCoDa \citep{CohenMcFarlane2020NovelCC}. Additionally, we conduct a comprehensive cross-datasets study (CDS), in which our proposed approach is first trained on one dataset and then evaluated on multiple other datasets. We then repeat this process by individually training the model on each remaining dataset and assessing its performance across the other datasets. Furthermore, all five datasets are consolidated into a combined dataset, which serves as the basis for the COVID-19 classification. We conducted extensive experiments on this unified dataset to evaluate the effectiveness of our proposed method, thoroughly evaluating its performance and robustness across diverse data sources to ensure comprehensive validation of our approach. The key contributions of this paper are summarized as follows:
\begin{itemize}

\item We propose a pioneering method that uses deep neural decision trees and decision forests tailored to detect COVID-19 from cough sounds. This approach offers a unique perspective on leveraging machine learning techniques for early diagnosis.

\item We use an advanced feature dimension reduction technique to enhance prediction performance, including recursive feature elimination with cross-validation and the extra-trees classifier. By identifying and prioritizing the most informative features, we significantly boost the accuracy and reliability of our detection model.

\item We explore five distinct training strategies across various frameworks to optimize the efficacy of our detection model. Leveraging Bayesian optimization to select the optimal hyperparameters for Deep Neural Decision Tree and Forest models significantly enhances the accuracy and reliability of our detection model, ensuring robust performance across various scenarios.

\item To validate the effectiveness and generalizability of our proposed approach, we conduct extensive evaluations on five diverse cough datasets. This rigorous evaluation framework ensures the reliability of our findings and underscores the versatility of our approach across different data sources.

\item We empirically assess the performance of our proposed approach by benchmarking it against state-of-the-art models for distinguishing COVID-19 from non-COVID-19 cases. Through comprehensive comparative analyses, we demonstrate the superiority of our method in terms of accuracy and efficiency.

\item To assess the generalizability of our approach, we perform cross-datasets evaluations wherein the model is trained on one dataset and evaluated on four additional datasets. This rigorous evaluation paradigm ensures the robustness and versatility of our detection method across diverse data sources and scenarios.

\end{itemize}

The subsequent sections of this paper are structured as follows. Section \ref{sec:relatedwork} presents the related work, while Section \ref{sec:rearchquestions} outlines the research questions addressed in this study. Section \ref{sec:methodology} details the methodology and explains our proposed approach in-depth. Section \ref{sec:results} presents the experimental results. Finally, Section \ref{sec:conclusion} concludes the article, summarizing the key findings and directions for future research.

\section{Related work}
\label{sec:relatedwork}

Numerous researchers have concentrated on detecting COVID-19 using sounds, such as coughs, breath, voice, and speech~\citep{khanna2022diagnosing, aleixandre2022use, santosh2022systematic, lella2021literature, ijaz2022towards, hasan2023review, gomes2022comprehensive, abdeldayem2022viral, deshpande2022ai}. In this study, we focus on research related to cough sounds. Many researchers have invested significant efforts in developing datasets, including Cambridge \citep{10.1145/3394486.3412865}, Virufy \citep{Chaudhari2020VirufyGA}, COUGHVID \citep{Orlandic2020TheCC}, Coswara \citep{Sharma2020CoswaraA}, Novel Coronavirus Cough Database (NoCoCoDa) \citep{CohenMcFarlane2020NovelCC}, Cough against COVID \citep{Bagad2020CoughAC}, AI4COVID-19 \citep{IMRAN2020100378}, MIT-Covid-19 \citep{Subirana2020HiSD}, IATos \citep{Pizzo2021IATosAP}, Sarcos \citep{Pahar2020COVID19CC}, ComParE \citep{inproceedingsCom}, COVID-19 Cough \citep{Laguarta2020COVID19AI}, COVID-19 Sounds \citep{Xia2021COVID19SA}, and DiCOVA Challenge \citep{Muguli2021DiCOVACD}.
Next, we investigate studies exclusively focusing on using cough sounds to classify COVID-19. We explore these studies' methodologies, findings, and implications to gain a comprehensive understanding of the role and effectiveness of cough-based approaches in COVID-19 detection.

There are two classifications of COVID-19 datasets: publicly available and inaccessible. Publicly available datasets include Virufy \citep{Chaudhari2020VirufyGA}, COUGHVID \citep{Orlandic2020TheCC}, Coswara \citep{Sharma2020CoswaraA}, IATos \citep{Pizzo2021IATosAP}, and others. Datasets that are not publicly available include Cambridge \citep{10.1145/3394486.3412865}, Novel Coronavirus Cough Database (NoCoCoDa) \citep{CohenMcFarlane2020NovelCC}, Cough against COVID \citep{Bagad2020CoughAC}, AI4COVID-19 \citep{IMRAN2020100378}, MIT-Covid-19 \citep{Subirana2020HiSD}, Sarcos \citep{Pahar2020COVID19CC}, ComParE \citep{inproceedingsCom}, COVID-19 Cough \citep{Laguarta2020COVID19AI}, COVID-19 Sounds \citep{Xia2021COVID19SA}, DiCOVA Challenge \citep{Muguli2021DiCOVACD}, and others.

A comparison of our research and previous studies on COVID-19 identification using cough sound analysis is shown in Table \ref{tab:related_work}. In the merged dataset (M) and cross-datasets study (CDS), a '-' symbol indicates that only one dataset is used. We perform a comprehensive CDS, where we initially train our proposed method on a single dataset and then evaluate its performance on several other datasets. We repeat this procedure by training the model on each remaining dataset one by one and assessing its results on the other datasets.
Moreover, a '\ding{51}' signifies that methods such as merged dataset (M), cross-dataset study (CDS), feature selection (FS), hyper-parameter tuning (HT), and threshold moving (TM) are applied, while a '\ding{55}' indicates they are not implemented. It is worth noting that some researchers limited their analyses to specific datasets, such as Cambridge \citep{10.1145/3394486.3412865,chowdhury2021qucoughscope,ashby2022cough,akgun2021transfer}, Coswara \citep{article10,article11,anupam2021preliminary,benmalek2023cough,benmalek2022cough}, COUGHVID \citep{Orlandic2022ASA,awais2023optimized,NAJARAN2023200,Sunitha2022ACA,hamdi2022attention,ren2022acoustic,inbook,10205895,Fakhry2021VirufyAM,ayappan2023mayfly,meng2022detection,cesarelli2022covid,deivasikamani2022covid,esposito2021covid}, and Virufy \citep{article6,kapoor2022cough,nafiz2023automated,ISLAM2022100025,MelekManshouri2021IdentifyingCB,erdougan2021covid,islam2021early}. A few others use only their proprietary datasets \citep{IMRAN2020100378,9361107,zealouk2021analysis,9256562,nasab2023coronavirus,10.1145/3412841.3441943}. Furthermore, in addition to using well-known datasets like Cambridge, Coswara, COUGHVID, and Virufy, some research included the use of other publicly and non-publicly accessible datasets \citep{Bagad2020CoughAC,RAYAN2023323,Ayyavaraiah_Venkateswarlu_2023,Despotovi2021DetectionOC,9208795,trang2022covid,malviya2023long,zhang2021novel,yan2022convoluational,nguyen2023fruit,9986431,deshpande2021DiCOVA,chang2022ufrc,wullenweber2022coughlime,mouawad2021robust} to validate the robustness of their classification models.

Numerous studies have evaluated the effectiveness of their suggested techniques using several datasets. Of these, a few studies used two datasets \citep{aytekin2023covid19,s23114996,article8,app13126976,Xue2021ExploringSR,9781738,10.1007/s00521-021-06346-3,PAHAR2021104572,mohammed2021ensemble,rao2021covid,zhang2022robust,pavel2022evaluation,harvill2021classification,tawfik2022multi,rao2021deep,irawati2021classification,10.1145/3500931.3500968,mehta2023proposed,9401826}, while others evaluated from three \citep{Chaudhari2020VirufyGA,sobahi2022explainable,P2022ScreeningCB,article5,article4,9989855,He2022TFACLSTMNNNC,PAHAR2022105153,shen2023piecewise} or four datasets \citep{Ulukaya2023MSCCov19NetMD,article2,tena2022automated,chang2022covnet,haritaoglu2022using}. Nevertheless, our work is noteworthy because it is the only one that fully incorporates each of the five well-known COVID-19 cough datasets: Cambridge, Coswara, COUGHVID, Virufy, and Virufy merged with NoCoCoDa. Also, some researchers choose to take a combined strategy, combining two or more datasets to strengthen the validity of their research. Among these studies, several combined two datasets \citep{article8,10.1007/s00521-021-06346-3,mohammed2021ensemble,rao2021deep,Ulukaya2023MSCCov19NetMD}, while several investigated combining three datasets \citep{P2022ScreeningCB,9989855} or four datasets \citep{mehta2023proposed,article2,tena2022automated,haritaoglu2022using} for a more thorough analysis. It is imperative to emphasize that, apart from our research, no one else has attempted the integration of the five well-known COVID-19 cough datasets: Cambridge, Coswara, COUGHVID, Virufy, and Virufy merged with NoCoCoDa.

The methodology of the cross-datasets study involves sequentially training the proposed method on specific datasets, one dataset used for training at a time, and subsequently evaluating its effectiveness across diverse datasets, excluding those used during the training phase. Several studies have undertaken partial cross-datasets investigations, where the proposed method is trained on a specified dataset with subsequent assessment across either a singular dataset or a variety of datasets different from the original training dataset. The researchers in \citet{Ulukaya2023MSCCov19NetMD} trained their suggested technique using a combined dataset that included COUGHVID and Coswara. Following the training phase, they evaluated their method using the Virufy and NoCoCoDa datasets. In the study conducted by \citet{PAHAR2021104572}, the researchers developed their suggested approach using the Coswara dataset and then tested it on the Sarcos dataset. Accordingly, the Coswara dataset was used by \citet{zhang2022robust} and \citet{9401826} to design and train their respective methods, with the Virufy dataset being used for performance assessments thereafter. Similarly, \citet{nguyen2023fruit} developed their suggested model with a publicly available dataset and evaluated its performance with the AICovidVN dataset. The studies mentioned earlier do not use all their datasets for cross-datasets investigations. Instead, they choose either a single dataset or a combined dataset for training and one or two distinct datasets for testing. In contrast, our research stands out as it performs a distinctive cross-datasets analysis using all five datasets (Cambridge, Coswara, COUGHVID, Virufy, and Virufy with NoCoCoDa). Notably, we use one of the five datasets for training our proposed methods, while the remaining four are independently used to validate each method.

Audio features are classified into different types, encompassing time domain features like Zero Crossing Rate (ZCR), Energy, Amplitude based features, Root Mean Square (RMS) Energy, etc. Also, frequency domain features include Power Spectral Density (PSD), Spectral Bandwidth, Spectral Contrast, Spectral Centroid, Spectral Roll-Off, Spectral Kurtosis, Spectral Flux, Spectral Spread, Chromagram, Tonal Centroid, and others. Time-frequency representations involve Spectrogram, Mel-Spectrogram, Constant-Q Transform, Scattering Embeddings, etc. Cepstral domain features comprise Mel-frequency Cepstral Coefficient (MFCC), $\Delta$-MFCC, $\Delta^2$-MFCC, Inverted Mel-Frequency Cepstral Coefficients (IMFCCs), Gamma-tone Frequency Cepstral Coefficients (GFCC), Gamma-tone Cepstral Coefficients (GTCCs), and so on. Deep features include characteristics obtained through deep learning models, such as Convolutional Neural Networks (CNNs), YAMNet, VGGish, and similar architectures.

In the field of COVID-19 classification, studies use various audio features for COVID-19 classification. Certain studies \citep{cesarelli2022covid,islam2021early} focus solely on frequency domain features, providing useful insights into the different patterns found in this domain. On the other hand, some studies \citep{chowdhury2021qucoughscope,akgun2021transfer,article10,hamdi2022attention,meng2022detection,esposito2021covid,nafiz2023automated,erdougan2021covid,RAYAN2023323,yan2022convoluational,chang2022ufrc,wullenweber2022coughlime,aytekin2023covid19,article8,rao2021deep,9989855,shen2023piecewise,tena2022automated,chang2022covnet} focus primarily on time-frequency representations, uncovering the unique properties hidden in these representations. Other investigations \citep{ashby2022cough,benmalek2023cough,benmalek2022cough,article6,kapoor2022cough,9208795,malviya2023long,zhang2021novel,9986431,mouawad2021robust,10.1007/s00521-021-06346-3,harvill2021classification,10.1145/3500931.3500968,mehta2023proposed,He2022TFACLSTMNNNC,LELLA20221319} focus on cepstral domain features, shedding light on the pertinent information concealed within cepstral analyses. Furthermore, a subset of researchers opts for deep features, harnessing the power of deep learning techniques to uncover intricate patterns within the audio data \citep{10.1145/3394486.3412865,app13126976}. 

Many studies are adopting a comprehensive approach by integrating multiple types of audio features into their analyses. For instance, some studies \citep{Chaudhari2020VirufyGA,IMRAN2020100378,MelekManshouri2021IdentifyingCB,9361107,Ayyavaraiah_Venkateswarlu_2023,trang2022covid,nguyen2023fruit,deshpande2021DiCOVA,Xue2021ExploringSR,zhang2022robust,pavel2022evaluation,article5,gupta2022cough} incorporate two different types of audio features. In contrast, others use three different types of features \citep{10.1145/3394486.3412865,Bagad2020CoughAC,anupam2021preliminary,Orlandic2022ASA,awais2023optimized,ayappan2023mayfly,ISLAM2022100025,9256562,10.1145/3412841.3441943,app13126976,PAHAR2021104572,tawfik2022multi,irawati2021classification,9401826,article4,PAHAR2022105153,Ulukaya2023MSCCov19NetMD,article2,haritaoglu2022using}. Notably, some studies use a combination of four distinct types of audio features ~\citep{10205895,nasab2023coronavirus,mohammed2021ensemble,rao2021covid,P2022ScreeningCB}. In our study, we used three distinct feature categories. These were features derived from the frequency domain, such as tonal centroid, chromagram, and spectral contrast; features from the cepstral domain, such as Mel-Frequency Cepstral Coefficients (MFCC); and features from the time-frequency representation domain, such as the Mel-Scaled Spectrogram. This approach exemplifies the detailed examination of numerous feature domains in the pursuit of robust COVID-19 classification models.

The technique of Optimal Feature Selection has been used in numerous research investigations. To choose the best features for training classification models, some studies \citep{Orlandic2022ASA,P2022ScreeningCB,article2,tena2022automated} used the Recursive Feature Elimination with Cross-Validation (RFECV) technique. To determine the optimal feature set, the Sequential Forward Selection (SFS) technique has been used in a number of studies \citep{MelekManshouri2021IdentifyingCB,10.1007/s00521-021-06346-3,PAHAR2021104572,PAHAR2022105153}. A few studies \citep{benmalek2023cough,erdougan2021covid} selected the best features using the Relief feature selection approach. A variety of feature selection techniques, including the Modified Cat and Mouse Based Optimizer (MCMBO), Stacked autoencoder architecture, Kruskal-Wallis test, Mutual information criterion, principal component analysis (PCA), and GridSearch, were used in a number of other studies \citep{Chaudhari2020VirufyGA,benmalek2022cough,awais2023optimized,inbook,Ayyavaraiah_Venkateswarlu_2023,Despotovi2021DetectionOC,mohammed2021ensemble} to optimize their feature sets. This work uses an Extra-Trees classifier with Recursive Feature Elimination with Cross-Validation (RFECV) to perform optimal feature selection. The choice to use RFECV is substantiated by the results presented in a study \citep{misra2020improving} that illustrated improved classification accuracy through RFECV.

Some researchers have performed hyper-parameter tuning to determine which parameters are optimal for their classification models. Numerous studies \citep{Chaudhari2020VirufyGA,ashby2022cough,akgun2021transfer,hamdi2022attention,inbook,deivasikamani2022covid,esposito2021covid,article6,Despotovi2021DetectionOC,rao2021covid,pavel2022evaluation,Ulukaya2023MSCCov19NetMD,article2} used Grid Search for hyper-parameter tuning. Optimal hyper-parameters for their suggested models are selected by a cross-validation procedure in a few studies \citep{Bagad2020CoughAC,IMRAN2020100378,aytekin2023covid19,Xue2021ExploringSR}. The optimal hyper-parameters were chosen using five-fold cross-validation in two studies \citep{ren2022acoustic,Sdergren2021DetectingCF}. Nested cross-validation has been used in certain studies \citep{10.1145/3394486.3412865,PAHAR2022105153} as a technique for hyper-parameter tuning in their analyses. Hyper-parameter optimization was done using a variety of methods in some other studies \citep{Orlandic2022ASA,awais2023optimized,ayappan2023mayfly,10.1007/s00521-021-06346-3,PAHAR2021104572}. These methods included MCMBO (Modified Cat and Mouse Based Optimizer), LOOCV (Leave-One-Out Cross-Validation), the Leave-p-out cross-validation, Leave-p-out nested cross-validation, and Mayfly optimization (MFO). Bayesian Optimization (BO) is more effective than more brute-force methods like Grid Search (GS) and Random Search (RS) in determining the ideal hyper-parameter combination \citep{eggensperger2013towards}.

Techniques for shifting thresholds have been used in various studies. Threshold moving based on accuracy scores was used in the study by \citet{hamdi2022attention}. ROC-AUC values were used to determine threshold movement in \citep{article2}. \citet{Bagad2020CoughAC} chose a threshold based on which the maximum Sensitivity score was obtained. In their investigations, \citet{zhang2021novel} and \citet{mouawad2021robust} used threshold moving based on the F1 score. Our study uses the threshold moving technique based on ROC-AUC values, as it strikes a balance between precision and recall.

Unlike prior investigations, our study thoroughly assesses model performance using cross-datasets study, a facet overlooked in previous research efforts. Also, we extend our analysis beyond the traditional scope of testing model performance across diverse datasets. Delving into the intricacies of training strategies, we scrutinize how different training methodologies significantly influence the overall performance of our models. This comprehensive examination ensures a holistic understanding of our model's capabilities, emphasizing the importance of both testing across varied datasets and adopting nuanced training strategies for robust and informed outcomes.


\LTcapwidth=\textwidth
{\footnotesize
\begin{longtable}{llcclccc}
\caption{A comparison to previous studies on COVID-19 detection from cough sounds.}\label{tab:related_work}\\ 
    
\hline
\multirow{2}{*}{Reference}                           & \multicolumn{3}{c}{Dataset}                                                                                                                                                                                                                                                       & \makecell[c]{\multirow{2}{*}{Feature extraction}}                                                          & \multicolumn{3}{c}{Approach}                                                                                                                                                                                                                                 \\ \cline{2-4} \cline{6-8} 
                                                               & \multicolumn{1}{l}{Name}                                                                                                                       & M & CDS                                                                                   &                                                                                                       & FS                                                                         & HT & TM           \\ \hline\hline
    \endfirsthead
    
    \multicolumn{3}{c}%
    {\tablename\ \thetable\ -- \textit{(Continued .....)}} \\
    
    \hline
\multirow{2}{*}{Reference}                           & \multicolumn{3}{c}{Dataset}                                                                                                                                                                                                                                                      & \makecell[c]{\multirow{2}{*}{Feature extraction}}                                                         & \multicolumn{3}{c}{Approach}                                                                                                                                                                                                                                 \\ \cline{2-4} \cline{6-8} 
                                                               & \multicolumn{1}{l}{Name}                                                                                                                       & M & CDS                                                                                   &                                                                                                       & FS                                                                         & HT & TM           \\ \hline
    \endhead
    
    \hline \multicolumn{4}{r}{} \\
    \endfoot
    
    \hline
    \endlastfoot

\citet{10.1145/3394486.3412865}  &     Cambridge                                                                        & - & - & \makecell[t l]{MFCC, $\Delta$-MFCC, $\Delta^2$-MFCC, RMS Energy,\\ Spectral Centroid, Roll-Off Frequency.}     & \ding{55}                                                      & \ding{51}                                                     & \ding{55}                                           \\

\citet{Chaudhari2020VirufyGA} &   \makecell[t l]{Coswara, COUGHVID,\\Virufy}                & \ding{55} & \ding{55} & MFCCs, Mel-Spectrogram    & \ding{51}                                                       & \ding{51}                                                     & \ding{55}                                           \\ 

\citet{Bagad2020CoughAC} &      \makecell[t l]{Cough against \\COVID}             & - & - & \makecell[t l]{MFCC, Mel-Spectrogram, RMS Energy, Tempo}     & \ding{55}                                                       & \ding{51}                                                     & \ding{51}                                           \\ 
\citet{IMRAN2020100378}  &     Own dataset                                                                  & - & - & MFCC, Mel-Spectrogram      & \ding{55}                                                       & \ding{51}                                                     & \ding{55}                                           \\ 
\citet{chowdhury2021qucoughscope} &   Cambridge                                                                         & - & - & Spectrogram & \ding{55}                                                       & \ding{55}                                                     & \ding{55}                                           \\ 
\citet{ashby2022cough}     &   Cambridge                                                                         & - & - & MFCC      & \ding{55}                                                       & \ding{51}                                                     & \ding{55}                                           \\ 
\citet{akgun2021transfer}   &    Cambridge                                                                    & - & - & \makecell[t l]{Mel-Spectrograms}    & \ding{55}                                                       & \ding{51}                                                     & \ding{55}                                           \\ 
\citet{article10}   &     Coswara                                                                        & - & - & Mel Scale Spectrogram                 & \ding{55}                                                       & \ding{55}                                                     & \ding{55}                                           \\ 

\citet{article11}   &     Coswara                                                                          & - & - & Genetic Algorithm (GA)                  & \ding{55}                                                       & \ding{55}                                                     & \ding{55}                                           \\ 
\citet{anupam2021preliminary}   &  Coswara                                                                      & - & - & \makecell[t l]{MFCCs, Spectral Centroid, Spectral Roll-Off Point,\\ Spectral Kurtosis, ZCR}   & \ding{55}                                                       & \ding{51}                                                     & \ding{55}                                           \\ 

\citet{benmalek2023cough}  &     Coswara                                                                     & - & - & \makecell[t l]{MFCC, Gamma-tone Cepstral Coefficients (GTCC)}    & \ding{51}                                                      & \ding{55}                                                     & \ding{55}                                           \\ 
\citet{benmalek2022cough} &  Coswara                                                                      & - & - & MFCC        & \ding{51}                                                       & \ding{55}                                                     & \ding{55}                                           \\ 
\citet{Orlandic2022ASA}    &    COUGHVID                                                                     & - & - &  \makecell[t l]{Power Spectral Density (PSD), Zero Crossing Rate \\(ZCR), Mel-frequency Cepstral Coefficients \\(MFCCs), etc.}         & \ding{51}         & \ding{51}                                                    & \ding{55}                                           \\ 
\citet{awais2023optimized}  &   COUGHVID                                                                      & - & - & \makecell[t l]{MFCC, Spectral and Statistical Features}        & \ding{51}                                                       & \ding{51}                                                     & \ding{55}                                           \\ 
\citet{NAJARAN2023200}  &       COUGHVID                                                                       & - & - & \makecell[t l]{OpenSMILLE, CNN, Statistical Feature}           & \ding{55}                                                      & \ding{55}                                                     & \ding{55}                                           \\ 

\citet{Sunitha2022ACA} &       COUGHVID                                                                       & - & - & CNN            & \ding{55}                                                      & \ding{55}                                                     & \ding{55}                                           \\ 

\citet{hamdi2022attention} &    COUGHVID                                                                       & - & - & Mel-Spectrogram            & \ding{55}                                                       & \ding{51}                                                     & \ding{51}                                           \\ 

\citet{ren2022acoustic}  &          COUGHVID                                                                         & - & - & OpenSMILE        & \ding{55}                                                       & \ding{51}                                                     & \ding{55}                                           \\ 

\citet{inbook} &         COUGHVID                                                                         & - & - & OpenSMILE                   & \ding{51}                                                       & \ding{51}                                                     & \ding{55}                                          \\ 

\citet{10205895}  &      COUGHVID                                                                         & - & - & \makecell[t l]{ZCR, MFCC, Chroma STFT, Roll-Off, Spectral \\Centroid, Spectral Bandwidth}                   & \ding{55}                                                       & \ding{55}                                                     & \ding{55}                                           \\ 

\citet{Fakhry2021VirufyAM} &     COUGHVID                                                                          & - & - & \makecell[t l]{MFCC, Mel-Spectrogram, Clinical Features}    & \ding{55}                                                       & \ding{55}                                                     & \ding{55}                                           \\ 

\citet{ayappan2023mayfly}   &   COUGHVID                                                                          & - & - & \makecell[t l]{MFCC, Log Frame Energies, ZCR, Kurtosis}     & \ding{55}                                                       & \ding{51}                                                     & \ding{55}                                           \\ 

\citet{meng2022detection}    &    COUGHVID                                                                         & - & - & Mel-Spectrogram   & \ding{55}                                                     & \ding{55}                                                     & \ding{55}                                           \\ 

\citet{cesarelli2022covid} &  COUGHVID                                                                     & - & - & Spectral Roll-Off      & \ding{55}                                                       & \ding{55}                                                     & \ding{55}                                           \\ 

\citet{esposito2021covid}   &   COUGHVID                                                                     & - & - & Log Mel Spectrograms     & \ding{55}                                                       & \ding{51}                                                     & \ding{55}                                           \\ 
\citet{article6}  &    Virufy                                                                         & - & - & MFCCs                    & \ding{55}                                                       & \ding{51}                                                     & \ding{55}                                           \\ 

\citet{kapoor2022cough} &      Virufy                                                                         & - & - & MFCC           & \ding{55}                                                       & \ding{55}                                                     & \ding{55}                                           \\ 
\citet{nafiz2023automated} &       Virufy                                                                         & - & - & Mel-Spectrogram        & \ding{55}                                                       & \ding{55}                                                     & \ding{55}                                           \\ 
\citet{ISLAM2022100025} &    Virufy                                                                         & - & - & \makecell[t l]{Time and Frequency Domain Features}              & \ding{55}                                                       & \ding{55}                                                     & \ding{55}                                           \\ 
\citet{MelekManshouri2021IdentifyingCB} &    Virufy                                                                         & - & - & \makecell[t l]{MFCC, Short-Time Fourier Transform (STFT)}     & \ding{51}                                                       & \ding{55}                                                     & \ding{55}                                           \\ 

\citet{erdougan2021covid}    & Virufy                                                                       & - & - & \makecell[t l]{Features on Intrinsic Mode Functions (IMFs)}     &  \ding{51}                                                       & \ding{55}                                                     & \ding{55}                                           \\ 

\citet{islam2021early}     &  Virufy                                                                       & - & - & Chromagram       & \ding{55}                                                       & \ding{55}                                                     & \ding{55}                                           \\ 
\citet{9361107}  &          Own dataset                                                        & - & - & \makecell[t l]{Mel-Scaled Spectrogram, Linear Predictive Coding \\Spectrum (LPCS), MFCC}                & \ding{55}                                                       & \ding{55}                                                     & \ding{55}                                           \\ 

\citet{zealouk2021analysis} &       Own dataset                                                        & - & - & \makecell[t l]{Hidden Markov Model (HMM), Gaussian Mixture\\ Distributions (GMMs), MFCC}        & \ding{55}                                                       & \ding{55}                                                     & \ding{55}                                           \\ 

\citet{9256562}  &      Own dataset                                                        & - & - & \makecell[t l]{MFCC, $\Delta$-MFCC, $\Delta^2$-MFCC, ZCR, Spectral Centroid,\\ Spectral Roll-Off}                    & \ding{55}                                                        & \ding{55}                                                     & \ding{55}                                           \\ 

\citet{nasab2023coronavirus} &    Own dataset                                                                  & - & - & \makecell[t l]{MFCC, $\Delta$-MFCC, $\Delta^2$-MFCC, Chromagram, etc.}   & \ding{55}                                                       & \ding{55}                                                    & \ding{55}                                           \\ 

\citet{10.1145/3412841.3441943} &   Own dataset                                                                  & - & - & MFCC, ZCR, Kurtosis, etc.   & \ding{55}                                                       & \ding{55}                                                     & \ding{55}                                           \\ 
\citet{RAYAN2023323}   &       MIT-Covid-19                                                                  & - & - & Mel-Spectrogram            & \ding{55}                                                       & \ding{55}                                                     & \ding{55}                                           \\ 

\citet{Despotovi2021DetectionOC}  &    CDCVA                                                                            & - & - & \makecell[t l]{GeMaps, eGeMaps, ComParE}     & \ding{51}                                                       & \ding{51}                                                    & \ding{55}                                           \\ 

\citet{9208795}   &      COVID-19 Cough                                                                    & - & - & MFCC            & \ding{55}                                                       & \ding{55}                                                     & \ding{55}                                           \\ 

\citet{trang2022covid}   &    AICV115M                                                                          & - & - & \makecell[t l]{MFCC, $\Delta$-MFCC, $\Delta^2$-MFCC, Log Frame Energies}       & \ding{55}                                                       & \ding{55}                                                     & \ding{55}                                           \\ 

\citet{malviya2023long}      &   \makecell[t l]{Pfizer digital medicine\\ challenge} & - & - & MFCC      & \ding{55}                                                       & \ding{51}                                                     & \ding{55}                                           \\ 

\citet{zhang2021novel}     &    Biovitals                                                                    & - & - & MFCCs     & \ding{55}                                                       & \ding{55}                                                     & \ding{51}                                           \\ 

\citet{yan2022convoluational} &    ComParE                                                                      & - & - & \makecell[t l]{3-channel Log mel Spectrograms}  & \ding{55}                                                      & \ding{51}                                                     & \ding{55}                                           \\ 

\citet{nguyen2023fruit} &   AICovidVN                                                                   & - & - & \makecell[t l]{Log Mel Spectrograms, Wavegram-Log Mel-CNN}         & \ding{55}                                                       & \ding{55}                                                     & \ding{55}                                           \\ 

\citet{9986431}    &    DiCOVA                                                                       & - & - & MFCC, $\Delta$-MFCC, $\Delta^2$-MFCC             & \ding{55}                                                       & \ding{55}                                                     & \ding{55}                                           \\ 
\citet{chang2022ufrc}   &     DiCOVA                                                                                    & - & - & Mel Spectogram      & \ding{55}                                                       & \ding{55}                                                     & \ding{55}                                           \\ 
\citet{wullenweber2022coughlime} &  DiCOVA                                                                                    & - & - & Spectrogram  & \ding{55}                                                       & \ding{55}                                                     & \ding{55}                                           \\ 
\citet{mouawad2021robust}   &    \makecell[t l]{Voca.ai and Carnegie\\Mellon University} & - & - & MFCC    & \ding{55}                                                       & \ding{55}                                                    & \ding{51}                                           \\ 
\citet{aytekin2023covid19} &        Cambridge, COUGHVID                                                             & \ding{55} & \ding{55} & Mel-Spectrogram       & \ding{55}                                                      & \ding{51}                                                     & \ding{55}                                           \\ 

\citet{s23114996} &       \makecell[t l]{COUGHVID,\\COVID-19 Sounds}            & \ding{55} & \ding{55} & \makecell[t l]{YAMNet, MFCC, VGGish, x-Vecs}                 & \ding{55}                                                       & \ding{55}                                                     & \ding{55}                                           \\ 

\citet{article8}    &      Cambridge, Qatari                                                               & \ding{51} & \ding{55} & Spectrogram                & \ding{55}                                                       & \ding{55}                                                     & \ding{55}                                           \\ 
\citet{app13126976} &       AICovidVN, Cambridge                                                            & \ding{55} & \ding{55} & \makecell[t l]{Mel Spectrum, MFCC, VGG Embeddings}               & \ding{55}                                                      & \ding{55}                                                     & \ding{55}                                           \\ 
\citet{Xue2021ExploringSR} &           \makecell[t l]{Coswara,\\COVID-19 Sounds}                                                         & \ding{55} & \ding{55} & \makecell[t l]{MFCC, Log compressed mel-filterbank}          & \ding{55}                                                      & \ding{51}                                                     & \ding{55}                                           \\ 
\citet{9781738}  &            Virufy, Cambridge                                                                & \ding{55} & \ding{55} & \makecell[t l]{Deep Wavelet Scattering Network (DWSN)}              & \ding{55}                                                       & \ding{55}                                                     & \ding{55}                                          \\ 

\citet{10.1007/s00521-021-06346-3} &    Virufy, NoCoCoDa                                                                 & \ding{51} & \ding{55} & MFCC    & \ding{51}                                                       & \ding{51}                                                     & \ding{55}                                           \\ 

\citet{PAHAR2021104572}  &         Coswara, Sarcos                        & \ding{55} & \ding{55} & \makecell[t l]{MFCC, $\Delta$-MFCC, $\Delta^2$-MFCC, etc.}         & \ding{51}                                                       & \ding{51}                                                     & \ding{55}                                           \\ 
\citet{mohammed2021ensemble} &     Coswara, Virufy                                                                            & \ding{51} & \ding{55} & \makecell[t l]{MFCC, Spectrogram, Power Spectrum, Chroma, etc.}     & \ding{51}                                                       & \ding{55}                                                     & \ding{55}                                           \\ 

\citet{rao2021covid}   &    DiCOVA, COUGHVID                                                                   & \ding{55} & \ding{55} & \makecell[t l]{MFCC $\Delta$-MFCC, $\Delta^2$-MFCC, RMS energy, etc.}         & \ding{55}                                                       & \ding{51}                                                     & \ding{55}                                           \\ 

\citet{zhang2022robust}   &   Coswara, Virufy                                                                   & \ding{55} & \ding{55} & \makecell[t l]{Log Mel Spectrograms, Time-frequency Differential \\Feature, Energy Ratio Feature}      & \ding{55}                                                       & \ding{55}                                                     & \ding{55}                                           \\ 

\citet{pavel2022evaluation}  &   COUGHVID, IATos                                                                    & \ding{55} & \ding{55} & MFCC, Mel-Spectrogram    & \ding{55}                                                       & \ding{51}                                                     & \ding{55}                                           \\ 

\citet{harvill2021classification} &  COUGHVID, DiCOVA                                                             & \ding{55} & \ding{55} & \makecell[t l]{MFCC, $\Delta$-MFCC, $\Delta^2$-MFCC}      & \ding{55}                                                      & \ding{55}                                                     & \ding{55}                                           \\ 

\citet{tawfik2022multi}    &  Coswara, Virufy                                                               & \ding{55} & \ding{55} & \makecell[t l]{MFCC, Chroma, ZCR, Spectral Centroid, Spectral \\Roll-Off, Spectral Bandwidth, Constant-Q Transform}       & \ding{55}                                                       & \ding{55}                                                     & \ding{55}                                           \\ 

\citet{rao2021deep}  &     DiCOVA, COUGHVID                                                              & \ding{51} & \ding{55} & Spectrograms         & \ding{55}                                                       & \ding{51}                                                     & \ding{55}                                           \\ 

\citet{irawati2021classification} &  Virufy, Coswara                                                               & \ding{55} & \ding{55} & \makecell[t l]{MFCC, Chroma, ZCR, Spectral Centroid, Spectral 
\\Bandwidth, Spectral Roll-Off, Root Mean Square}   & \ding{51}                                                       & \ding{51}                                                     & \ding{55}                                           \\ 

\citet{10.1145/3500931.3500968} &  \makecell[t l]{DiCOVA, Own dataset}           & \ding{55} & \ding{55} & MFCC, $\Delta$-MFCC, $\Delta^2$-MFCC   & \ding{55}                                                       & \ding{55}                                                     & \ding{55}                                           \\ 

\citet{mehta2023proposed}    &    \makecell[t l]{Coswara, COUGHVID,\\Virufy, Owndataset}    & \ding{51} & \ding{55} & MFCC  & \ding{55}                                                       & \ding{55}                                                     & \ding{55}                                           \\ 

\citet{9401826}    &      Coswara, Virufy                                                                            & \ding{55} & \ding{55} & \makecell[t l]{MFCC, Energy, Entropy of Energy, ZCR, Spectral\\ Centroid, Spectral Spread, Spectral Entropy, \\Spectral Flux}          & \ding{55}                                                       & \ding{51}                                                     & \ding{55}                                           \\ 
\citet{sobahi2022explainable} &      \makecell[t l]{COUGHVID, Virufy,\\Coswara}            & \ding{55} & \ding{55} & Fractal Dimensions (FD)      &\ding{55}                                                       & \ding{55}                                                     & \ding{55}                                          \\ 
\citet{P2022ScreeningCB} &       \makecell[t l]{Own dataset,\\ COUGHVID, Cambridge} & \ding{51} & \ding{55} & \makecell[t l]{MFCC, $\Delta$-MFCC, $\Delta^2$-MFCC, etc.}         & \ding{51}                                                       & \ding{55}                                                     & \ding{55}                                           \\ 

\citet{article5} &         \makecell[t l]{Cambridge, Coswara,\\ COUGHVID}          & \ding{55} & \ding{55} & \makecell[t l]{MFCC, $\Delta$-MFCC, $\Delta^2$-MFCC, Spectral Contrast}                & \ding{55}                                                       & \ding{55}                                                     & \ding{55}                                           \\ 

\citet{article4}  &          \makecell[t l]{COUGHVID, Cambridge, \\ Coswara}         & \ding{55} & \ding{55} & \makecell[t l]{MFCC, Spectrogram, Spectral Centroid, etc.}              & \ding{55}                                                       & \ding{55}                                                     & \ding{55}                                           \\ 

\citet{9989855}  &    \makecell[t l]{Own dataset,\\ Coswara, COUGHVID}   & \ding{51} & \ding{55} & Mel-Spectrogram                     & \ding{55}                                                       & \ding{55}                                                     & \ding{55}                                           \\ 

\citet{He2022TFACLSTMNNNC} &      \makecell[t l]{China,Virufy(Clinical),\\ Virufy(India)} & \ding{55} & \ding{55} & MFCC          & \ding{55}                                                       & \ding{55}                                                     & \ding{55}                                           \\ 

\citet{PAHAR2022105153} &      \makecell[t l]{Coswara, ComParE,\\Sarcos}               & \ding{55} & \ding{55} & \makecell[t l]{MFCC, $\Delta$-MFCC, $\Delta^2$-MFCC, etc.}             & \ding{51}                                                       & \ding{51}                                                     & \ding{55}                                           \\ 

\citet{shen2023piecewise}   &   \makecell[t l]{ComParE, Coswara,\\Cambridge}              & \ding{55} & \ding{55} & \makecell[t l]{Log Mel Spectrograms,Time-frequency Differential \\Feature}     & \ding{55}                                                       & \ding{55}                                                     & \ding{55}                                           \\ 
\citet{Ulukaya2023MSCCov19NetMD} &    \makecell[t l]{COUGHVID, Coswara,\\Virufy, NoCoCoDa}   & \ding{51} & \ding{55} & \makecell[t l]{MFCC, Spectrogram, Chromagram}     & \ding{55}                                                       & \ding{51}                                                     & \ding{55}                                           \\ 

\citet{article2} &       \makecell[t l]{Cambridge, Coswara\\Virufy, NoCoCoDa}  & \ding{51} & \ding{55} & \makecell[t l]{MFCC, Chromagram, Tonal Centroid, Spectral \\Contrast, Mel-Scaled Spectrogram}                  &  \ding{51}                                                       & \ding{51}                                                     & \ding{51}                                           \\ 

\citet{tena2022automated} &    \makecell[t l]{Lleida, Cambridge,\\Coswara,Virufy}        & \ding{51} & \ding{55} & Time-frequency Features      & \ding{51}                                                       & \ding{55}                                                     & \ding{55}                                           \\ 

\citet{chang2022covnet}      &   \makecell[t l]{Flusense, COUGHVID,\\ ComParE, DiCOVA}  & \ding{55} & \ding{55} & Log Mel Spectrogram    & \ding{55}                                                       & \ding{55}                                                     & \ding{55}                                           \\ 

\citet{haritaoglu2022using}   &   \makecell[t l]{COUGHVID, Coswara,\\ Virufy, IATos}     & \ding{51} & \ding{55} & \makecell[t l]{MFCC, $\Delta$-MFCC, $\Delta^2$-MFCC, Spectral Centroid, etc.}   & \ding{55}                                                       & \ding{51}                                                     & \ding{51}                                           \\ 
\citet{LELLA20221319} &      \makecell[t l]{COVID-19 Sound \\Crowdsourced Data}                                                                         & - & - & \makecell[t l]{De-noising Auto Encoder (DAE), Gamma-tone \\ Frequency Cepstral Coefficients (GFCC), Improved \\Multi-frequency Cepstral Coefficients (IMFCC)}        & \ding{55}                                                       & \ding{55}                                                     & \ding{55}                                           \\ 
\citet{gupta2022cough}   &     COUGHVID                                                                          & - & - & \makecell[t l]{MFCC, Mel-Frequency Spectrogram}      & \ding{55}                                                       & \ding{55}                                                     & \ding{55}                                           \\ 
\citet{Sdergren2021DetectingCF} &    DiCOVA                                                                          & - & - & OpenSMILE       & \ding{55}                                                       & \ding{51}                                                     & \ding{55}                                           \\ \hline

Our study    &            \makecell[t l]{Cambridge, Coswara,\\ COUGHVID, Virufy,\\NoCoCoDa} & \ding{51} & \ding{51} & \makecell[t l]{MFCC, Chromagram, Tonal Centroid, Spectral \\Contrast, Mel-Scaled Spectrogram}                        & \ding{51}                                                     & \ding{51}                                                     & \ding{51}                                           \\ \hline

\end{longtable} 
\vspace{-.7cm}
\noindent M - Merged, CDS - Cross-datasets study, FS - Feature selection, HT - Hyper-parameter tuning, TM - Threshold moving.
}

\section{Research questions}
\label{sec:rearchquestions}

In our pursuit of advancing the cutting-edge in cough-based COVID-19 detection, we present a ML-based architecture tailored to analyze cough sounds. Furthermore, our research focuses on pinpointing the most successful techniques for precise COVID-19 detection using cough data. As a result, we have devised a series of research questions (RQs) dedicated to the field of cough-based COVID-19 detection:

\begin{itemize}
\setlength{\itemsep}{1pt}

\item \textbf{RQ1:} How do different training strategies impact the classification performance of detecting COVID-19 from cough sounds?\\
\textemdash~We provide various training strategies to enhance the effectiveness of the proposed method. The significance of these training techniques is elucidated in Section~\ref{sec:trainingstrategies}, with a detailed analysis presented in Section~\ref{Classificationperformanceoffourstrategies}.

\item \textbf{RQ2:} Should the construction of detection models consider any demographic or geographic variations in cough sounds associated with COVID-19?\\
\textemdash~Yes, demographic and geographic variations in cough sounds related to COVID-19 exist. These variations should be considered when developing detection models to ensure they are effective and generalizable across different populations. We explore the specifics of this cross-datasets investigation and its resulting findings in Section~\ref{CrossSectionalStudy}.

\end{itemize}

\section{Methodology}
\label{sec:methodology}

\begin{figure}
\centering
  \includegraphics[width=\linewidth]{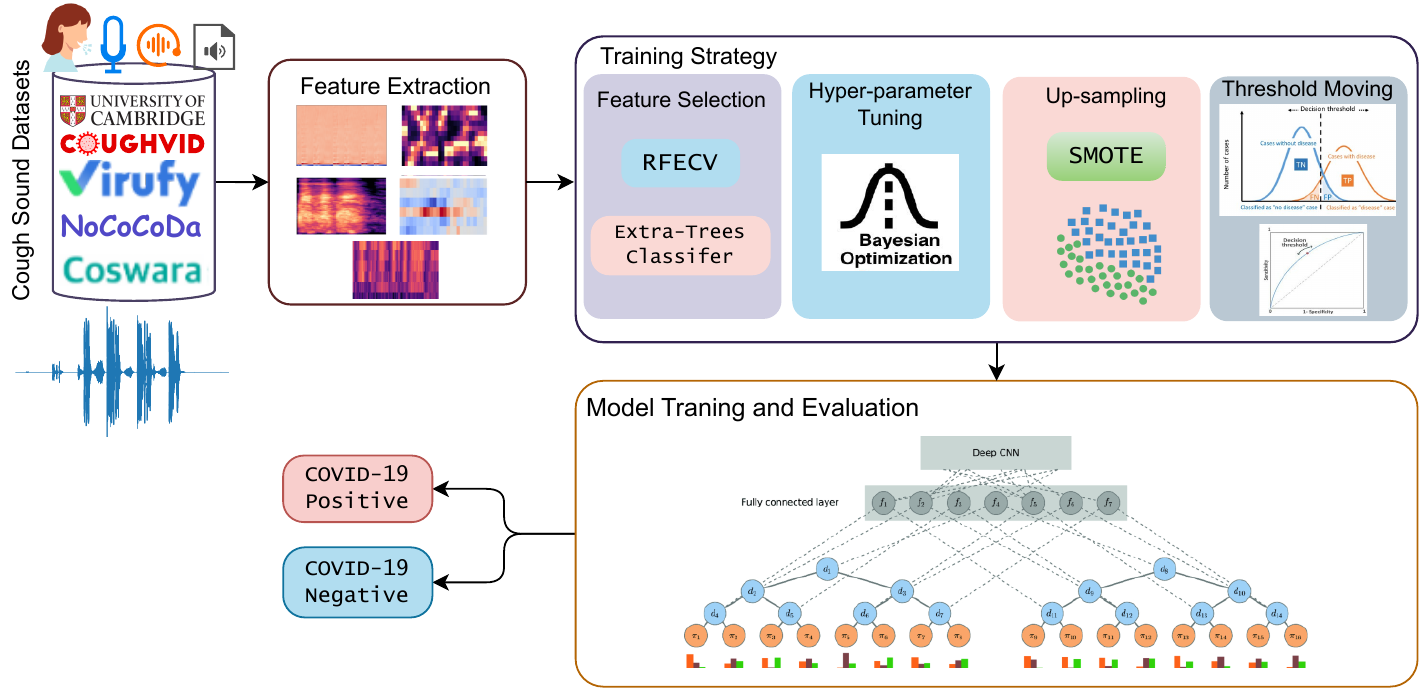}
  \caption{An overview of the proposed method to classify COVID-19 based on the analysis of cough sound audio recordings.}
  \label{fig:covid-19}
\end{figure}

Inspired by the advancements in ML-based audio applications, we have created a comprehensive ML framework capable of taking cough samples and making direct predictions of binary classification labels, hinting at the potential presence of COVID-19. Figure \ref{fig:covid-19} presents our methodology for COVID-19 detection from the cough sound. This method comprises several stages: data curation, feature extraction, feature selection and model evaluation. We used cough data samples from various crowdsource datasets during the data collection phase. The core of our proposed method relies on audio features, such as Mel-Frequency Cepstral Coefficients (MFCCs), Mel-Scaled Spectrogram, Tonal Centroid, Chromagram, and Spectral Contrast, followed by a feature fusion process. We use feature dimension reduction techniques, like RFECV and the Extra-Trees classifier, to identify the most important features. This feature representation and the hyper-parameter space are fed into a Bayesian Optimization function, which optimizes the hyper-parameters of our proposed classifiers. The Bayesian Optimization function then calculates the best hyper-parameters before training our proposed classifiers. The dataset
exhibits an under-representation of the positive category for COVID-19, which could potentially harm the ML classifier's performance. We have incorporated SMOTE during training to address this imbalance, aiming to balance the dataset and improve the ML classifier's performance. In classification tasks, using the default threshold (i.e., 0.50) often results in subpar performance, especially when dealing with class imbalance. As a remedy, we use the threshold moving technique to adapt the probability threshold that determines the assignment of class labels. The optimal threshold, hyper-parameters, and selected features are subsequently input into classifiers (DNDF and DNDT) for COVID-19 detection and to assess the proposed method. A comprehensive explanation of each phase in the proposed method is provided in the subsequent sections.

\subsection{Dataset description}
\label{sec:datasetdescription}

In this section, we describe the datasets we use to validate and evaluate the effectiveness of our COVID-19 classification models based on cough sound analysis. We use five different datasets in our experiment: Cambridge \citep{10.1145/3394486.3412865}, Coswara \citep{Sharma2020CoswaraA}, COUGHVID \citep{Orlandic2020TheCC}, Virufy \citep{Chaudhari2020VirufyGA}, and Virufy merged with NoCoCoDa \citep{CohenMcFarlane2020NovelCC}. Cambridge has a dataset of two different categories (asymptomatic and symptomatic). Every cough sample undergoes resampling with a sampling rate of 22.5 kHz, using a Hann window type. We also combined all five datasets to create a combined dataset with various cough samples from both COVID-19 positive and negative individuals, shown in Table \ref{tab:Table 1}.

\begin{table}[!t]
\centering
\caption{Cough samples from both COVID-19 positive and negative cases are used in separate datasets.}
\label{tab:Table 1}
\begin{tabular}{lcccc}
\hline
Dataset          & \multicolumn{1}{c}{Category} & COVID-19 & Non COVID-19 & Total \\ \hline \hline
\multirow{2}{*}{Cambridge \citep{10.1145/3394486.3412865}} & \multicolumn{1}{c}{Asymptomatic}       & 141               & 298                   & 439            \\ 
                           & \multicolumn{1}{c}{Symptomatic}        & 54                & 32                    & 86             \\                          
Coswara \citep{Sharma2020CoswaraA}                    & -                                      & 185               & 1,134                  & 1,319           \\ 
COUGHVID \citep{Orlandic2020TheCC}                   & -                                      & 680               & 680                   & 1,360           \\   
Virufy \citep{Chaudhari2020VirufyGA}                     & -                                      & 48                & 73                    & 121            \\ 
NoCoCoDa~\citep{CohenMcFarlane2020NovelCC}                   & -                                      & 73                & -                     & 73             \\ 
Virufy + NoCoCoDa \citep{CohenMcFarlane2020NovelCC}          & -                                      & 121               & 73                    & 194            \\ \hline
Combined          & - & 1,181               & 2,217                    & 3,398            \\ \hline

\hline
\end{tabular}
\end{table}

\subsubsection{Cambridge dataset}
The University of Cambridge has developed an online platform and mobile app that allows people to submit recordings of their coughs, inhalations, and voices while reciting a specific phrase. The Cambridge dataset~\citep{10.1145/3394486.3412865} is divided into two groups, asymptomatic and symptomatic, to distinguish between people who have tested positive for COVID-19 and those who have not. We acknowledge the limitations imposed by the authors of the Cambridge dataset, which is only available under a bilateral legal agreement for research purposes and not for commercial use.

\begin{itemize}
    \item \textbf{Asymptomatic:}
    To distinguish between people who have tested positive for COVID-19 and those who have tested negative, the Cambridge Asymptomatic dataset contains 141 cough samples from COVID-19 positive people and 298 cough samples from people who have tested negative for COVID-19. The people in the dataset have no notable medical conditions, do not smoke, and are asymptomatic (show no symptoms).
    \item \textbf{Symptomatic:} To distinguish between people who have tested positive for COVID-19 and people who have tested negative, both with a cough but no other medical conditions or smoking history, the Cambridge Symptomatic dataset contains 54 cough samples from COVID-19 positive people and 32 cough samples from people who have tested negative for COVID-19.
\end{itemize}

\subsubsection{Coswara dataset}
The Coswara dataset~\citep{Sharma2020CoswaraA} is a publicly available dataset of cough samples developed by the Coswara project, a collaboration between the Indian Institute of Science and the Indian Institute of Technology Palakkad. It was collected between April 2020 and May 2021 and contains 1,319 cough samples, with 185 samples from COVID-19 positive individuals and 1,134 samples from COVID-19 negative individuals. The samples were preprocessed and labeled based on the healthy (COVID-19 negative) and heavy cough variations (COVID-19 positive).  

\subsubsection{COUGHVID dataset}
The COUGHVID dataset~\citep{Orlandic2020TheCC} was collected by researchers at the Embedded System Laboratory (ESL) in Switzerland. We preprocessed and labeled the samples into two groups: those from healthy individuals (COVID-19 negative) and those from individuals with notable cough variations (COVID-19 positive). The COUGHVID dataset contains 1,360 cough samples, with 680 samples from COVID-19 positive individuals and 680 samples from COVID-19 negative individuals.

\subsubsection{Virufy dataset}
The Virufy COVID-19 open cough dataset~\citep{Chaudhari2020VirufyGA} is the first publicly available dataset of cough sounds from COVID-19 patients. The sounds were recorded in a hospital with the patient's consent, under the supervision of a physician, and in accordance with standard operating procedures. The Virufy dataset contains 121 cough samples from 16 patients, with 48 samples from COVID-19 positive patients and 73 samples from COVID-19 negative patients.

\subsubsection{NoCoCoDa dataset}
The NoCoCoDa dataset~\citep{CohenMcFarlane2020NovelCC} is a collection of cough sounds from COVID-19 patients recorded during interviews and news programs. It contains 73 cough sounds from 10 participants who attended 13 interviews. To provide a more comprehensive dataset for experiments, we combine the NoCoCoDa dataset with the Virufy dataset, which contains cough samples from both COVID-19 positive and negative people. The combined dataset contains 194 samples, with 121 from COVID-19 positive people and 73 from COVID-19 negative people.

\subsubsection{Combined dataset}
We consolidate the Cambridge (both asymptomatic and symptomatic), Coswara, COUGHVID, Virufy, and NoCoCoDa datasets to form a combined dataset. This unified dataset comprises 3,398 samples, including 1,181 cough samples from COVID-19-positive individuals and 2,217 cough samples from COVID-19-negative individuals.

\subsection{Feature extraction}
\label{sec:featureextractionmethods}

\begin{figure}[!ht]
  \centering
  \includegraphics[width=\textwidth]{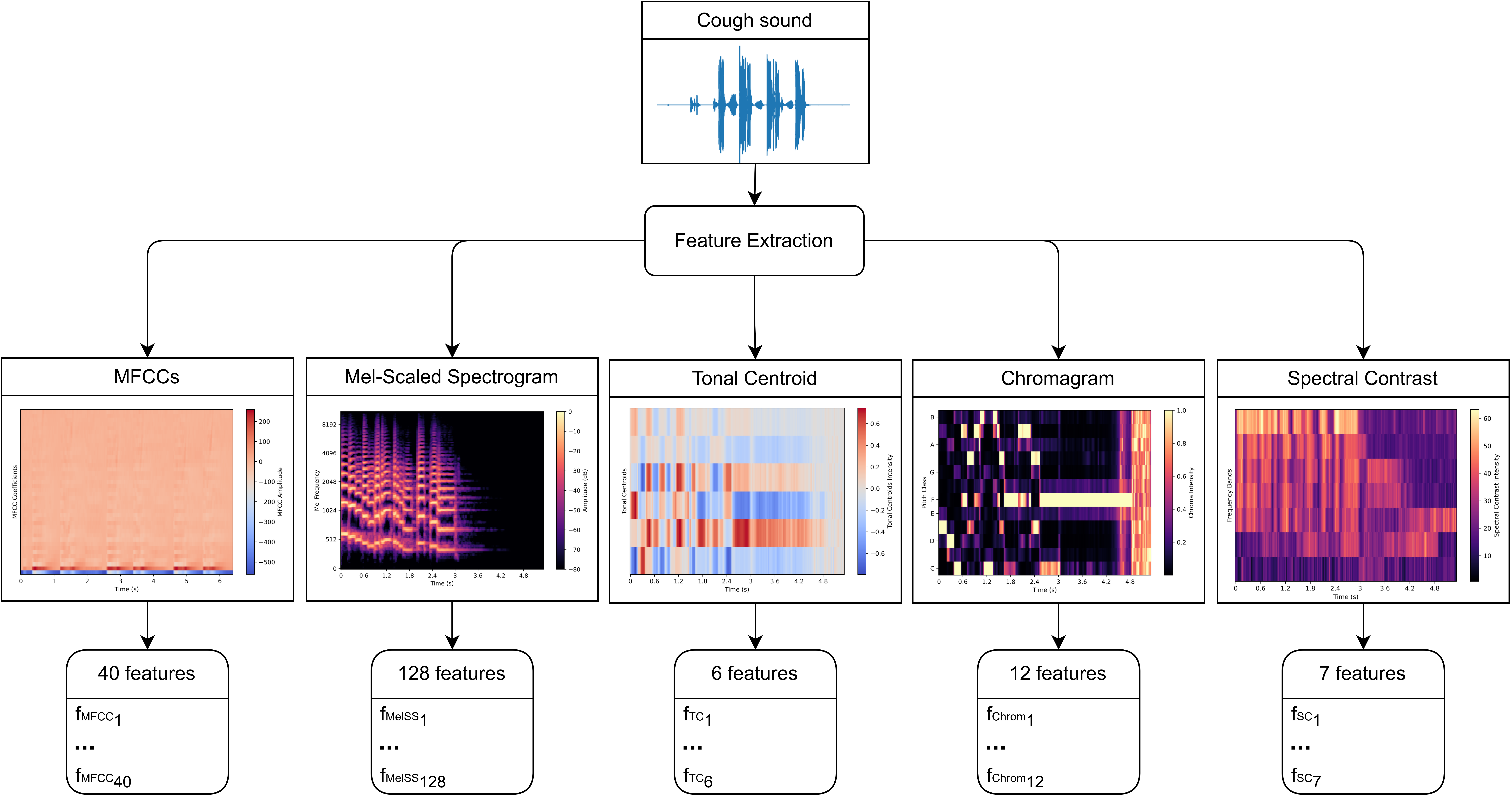}
  \caption{Overview of the extraction process for 193 features across five types: MFCCs, Mel-Scaled Spectrogram, Tonal Centroid, Chromagram, and Spectral Contrast.}\label{fig:features}
\end{figure}

To maintain consistency with a common standard in audio applications, we capture the acoustic signal used for feature extraction at a frequency of 22 kHz. We then compute five spectral feature types from the sampled audio: MFCCs, Mel-Scaled Spectrogram, Tonal centroid, Chromagram, and Spectral contrast. Figure \ref{fig:features} presents an overview of our feature extraction process. The Python library librosa \citep{brian_mcfee-proc-scipy-2015} is used to extract them.

\subsubsection{Mel-Frequency Cepstral Coefficients (MFCCs)}

MFCCs have demonstrated their effectiveness in distinguishing between dry and wet coughs \citep{5966670} and are well-regarded as valuable spectral features for audio analysis. From the audio signal, we extract 40 MFCC features. The process of extracting MFCCs encompasses several key steps. Initially, audio signals are divided into frames, each subject to a windowing function to mitigate noise from sudden changes at its start and end. Subsequently, the Fast Fourier Transform (FFT) is applied to compute the power spectrum of each frame post-windowing. This power spectrum is further manipulated using a filter bank design based on the Mel scale, as depicted in Equation \ref{equation1}, to obtain Mel-scaled filters from the original frequency (f). Ultimately, the Discrete Cosine Transform (DCT) is used to derive a set of MFCCs (MFCC coefficients) for each frame from the audio input, following the transformation of the power spectrum into a logarithmic scale. 

\begin{equation}
   \label{equation1}
    f_{mel} = 2595log_{10} (1+\dfrac{f}{700})
\end{equation}



\subsubsection{Mel-Scaled Spectrogram}
The Mel-Scaled Spectrogram is a widely adopted technique in ML for audio analysis, serving as a prevalent method for feature extraction from audio data. This process involves converting the power spectrogram into the Mel scale domain using a set of Mel filters. To generate a Mel-scaled Spectrogram, the initial step is to divide the signal into small frames using windowing. A window size of 2048 samples and a hop length of 512 samples are typically set for audio processing. Subsequently, the power spectrum is computed for each frame using the Fourier Transform. The number of Mel filters, usually set to 128, is evenly spaced in terms of frequencies on the Mel scale. Finally, the power spectrum is passed through these 128 Mel filters, followed by applying a logarithmic transformation to the resultant values. We obtain 128 Mel-scaled Spectrogram features from an audio signal. 

\subsubsection{Tonal centroid}
The tonal centroid, a feature used in audio analysis, is created by projecting a 12-bin chroma vector onto a six-dimensional vector using a transformation matrix, as denoted by Equation \ref{equation2} \citep{harte2006amcmm}. 

\begin{equation}
   \label{equation2}
    \zeta_{n} (d) = \dfrac{1}{\lvert \lvert  c_{n} \rvert  \rvert} \sum\limits_{l=0}^{11} \phi(d,l) c_{n}(l),\hspace{0.2cm}0 \le d < 5,\hspace{0.2cm}0 \le l \le 11
\end{equation}
where $\zeta_{n}$ is the 6-dimensional tonal centroid vector, computed by multiplying the transformation matrix $\Phi$ by the chroma vector c during the specified time frame n and dividing the resulting vector by the L1-norm of the chroma vector to ensure proper scaling of the values.

\subsubsection{Chromagram}
In the realm of ML applied to audio analysis, chromagrams serve as fundamental input features. We extracted 12 chromagram features from an audio signal. Generating a chromagram from an acoustic signal involves using the frequency power spectrum derived from the Short-Time Fourier Transform (STFT). The STFT is computed by using a sliding window over the audio signal and performing the Fourier transform for each window, effectively representing the audio stream as a time-frequency wave. Subsequently, the power spectrum is derived by squaring the magnitudes of the STFT coefficients.

The chromagram itself is derived from the power spectrum of an acoustic signal by mapping the frequency bins. In this context, a specific hop length of 512 and a window size of 2048 are chosen, creating 12 chroma bins. Finally, the feature vector is compiled by obtaining the normalized energy of each chroma bin for every frame in the audio signal.

\subsubsection{Spectral contrast}

Spectral contrast features find application in ML for audio analysis. From the audio signal, we extract seven spectral contrast features. The procedure for deriving spectral contrast features from an audio signal encompasses several sequential stages. Firstly, a Fast Fourier Transform (FFT) is applied to the digital audio clips, capturing the spectral distribution of the audio signal. Subsequently, the frequency spectrum is partitioned into a collection of sub-bands using octave band filters. The number of these sub-frequency bands is standardized at 6. The evaluation of spectral valleys, peaks, and their disparities is performed within each sub-band, as described in Equations \ref{equation3}, \ref{equation4} and \ref{equation5} \citep{Jiang2002MusicTC}. The initial spectral contrast values are then transformed into a logarithmic representation. Lastly, using a Karhunen-Loeve transform, the Log-frequency contrast values are projected into an orthogonal subspace.

\begin{equation} \label{equation3}
Peak_{k} = log\Big\{ \dfrac{1}{\alpha N} \sum\limits_{i=1}^{\alpha N} x_{k,i} \Big\}
\end{equation}

\begin{equation} \label{equation4}
Valley_{k} = log\Big\{ \dfrac{1}{\alpha N} \sum\limits_{i=1}^{\alpha N} x_{k,N-i+1} \Big\}
\end{equation}

\begin{equation} \label{equation5}
SC_{k} = Peak_{k} - Valley_{k}
\end{equation}
where N represents the overall count within the k-th sub-frequency band, k ranging from 1 to 6, and $\alpha$ is invariant with a range of 0.02 to 0.2.

\subsection{Feature selection}
\begin{figure}[!ht]
\centering
  \includegraphics[width=1\linewidth]{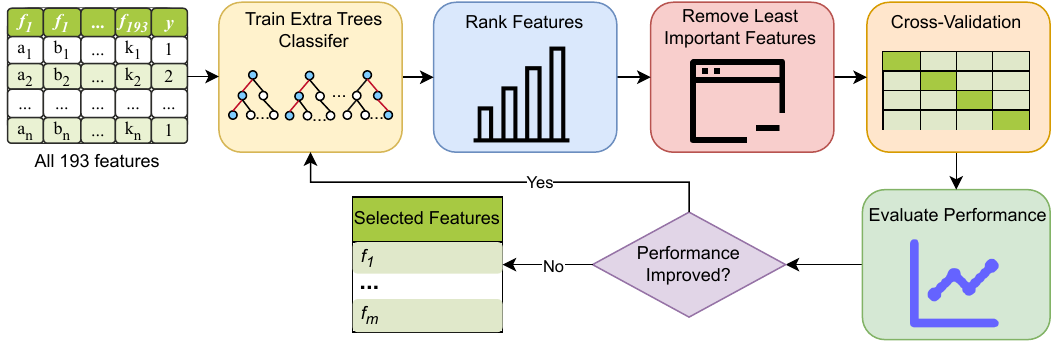}
  \caption{Feature selection with the recursive feature elimination with cross-validation (RFECV) and Extra-Trees classifier.}
  \label{fig:rfecv}
\end{figure}

We extract 193 features from each audio signal, 40 MFCCs, 128 mel-scaled spectrogram, 6 tonal centroid, 12 chromagram, and 7 spectral contrast features. However, it is worth noting that not all of these features are optimal. One method that is often used in ML is feature selection. Combining the Recursive Feature Elimination with Cross-Validation (RFECV) technique \citep{misra2020improving} and the Extra-Trees Classifier is an efficient method for feature dimensionality reduction (optimal feature selection). Figure~\ref{fig:rfecv} provides an overview of the RFECV method with the Extra Trees classifier for selecting optimal features. The most appropriate features to be chosen automatically are found using RFECV, which uses cross-validation and feature significance weights. Less significant characteristics are removed iteratively, and cross-validation is used to assess the model's performance. Following the use of the RFECV technique and Extra-Trees Classifier, we obtain optimal features of 71, 182, 33, 172, 46, and 188 for the Cambridge asymptomatic, Cambridge symptomatic, Coswara, COUGHVID, Virufy, and Virufy merged with NoCoCoDa datasets, respectively. We use the Extra-Trees estimator to obtain information about each feature’s significance. This method's main objective is to use the RFECV and Extra-Trees classifier to examine feature importance and minimize feature dimensions. 

\subsection{Hyper-parameter tuning}
Hyper-parameters control the learning process of an ML model, such as the number of trees, depth, used features rate, and epochs. Hyper-parameter tuning is finding the best values for these hyper-parameters to optimize the model's performance on a given dataset. Bayesian Optimization (BO) is a popular and effective technique for hyper-parameter tuning. It works by building a probabilistic model of the relationship between the hyper-parameters and the model's performance. It then uses this model to select the next set of hyper-parameters to find the values that lead to the best performance. BO uses a model to make informed decisions about which hyper-parameter values to test next, leveraging past results to make more efficient choices. This approach requires fewer iterations to find the optimal hyper-parameter combination than the more brute-force methods of grid search (GS) and random search (RS) \citep{eggensperger2013towards}. This efficiency is a key advantage of BO in hyper-parameter tuning. The performance of DNDT and DNDF for classification is significantly impacted by hyper-parameters. In BO, we define hyper-parameter space for hyper-parameters of our classifiers. Extracted input features and their labels are given to the Bayesian Optimization function to obtain the most effective hyper-parameter values. The default hyper-parameters used for all datasets in both DNDT and DNDF classifiers are presented in Table \ref{tab:table3}. Additionally, Table \ref{tab:table4} reports the optimized hyper-parameters for various datasets in both DNDT and DNDF classifiers.

\begin{table}[!ht]
\centering
\caption{The default hyper-parameters are used for all datasets.}
\label{tab:table3}
\begin{tabular}{cccccc}
\hline
\multicolumn{1}{c}{Num\_trees} & \multicolumn{1}{c}{Depth} & \multicolumn{1}{c}{Features rate} & \multicolumn{1}{c}{Learning rate} & \multicolumn{1}{c}{Batch size} & Num\_epochs \\ \hline \hline
\multicolumn{1}{c}{10}                  & \multicolumn{1}{c}{10}             & \multicolumn{1}{c}{1}                      & \multicolumn{1}{c}{0.01}                   & \multicolumn{1}{c}{256}                 & 1                   \\ \hline

\hline
\end{tabular}
\end{table}

\begin{table}[!ht]
\centering
\caption{The optimized hyper-parameters are used for different datasets.}
\label{tab:table4}
\resizebox{\textwidth}{!}{
\begin{tabular}{lccccccc}
\hline
\multirow{2}{*}{Dataset} & \multirow{2}{*}{Category} & \multicolumn{6}{c}{Hyper-parameter}                                                                                                                                                                                               \\ \cline{3-8} 
                                  &                                    & \multicolumn{1}{c}{Num\_trees} & \multicolumn{1}{c}{Depth} & \multicolumn{1}{c}{Features rate} & \multicolumn{1}{c}{Learning rate} & \multicolumn{1}{c}{Batch size} & Num\_epochs \\ \hline \hline
\multirow{2}{*}{Cambridge}        & Asymptomatic                        & \multicolumn{1}{c}{18}                  & \multicolumn{1}{c}{8}              & \multicolumn{1}{c}{0.8}                    & \multicolumn{1}{c}{0.01}                   & \multicolumn{1}{c}{32}                  & 17                   \\ & Symptomatic                         & \multicolumn{1}{c}{25}                  & \multicolumn{1}{c}{9}              & \multicolumn{1}{c}{0.8}                    & \multicolumn{1}{c}{0.01}                   & \multicolumn{1}{c}{8}                   & 13                   \\ 
Coswara                           & \multicolumn{1}{c}{-}             & \multicolumn{1}{c}{25}                  & \multicolumn{1}{c}{11}             & \multicolumn{1}{c}{0.6}                    & \multicolumn{1}{c}{0.01}                   & \multicolumn{1}{c}{16}                  & 14                   \\ 
COUGHVID                          & \multicolumn{1}{c}{-}             & \multicolumn{1}{c}{17}                  & \multicolumn{1}{c}{6}              & \multicolumn{1}{c}{1.0}                    & \multicolumn{1}{c}{0.01}                   & \multicolumn{1}{c}{32}                  & 19                   \\ 
Virufy                            & \multicolumn{1}{c}{-}             & \multicolumn{1}{c}{16}                  & \multicolumn{1}{c}{16}             & \multicolumn{1}{c}{0.8}                    & \multicolumn{1}{c}{0.01}                   & \multicolumn{1}{c}{8}                   & 22                   \\ 
Virufy + NoCoCoDa                 & \multicolumn{1}{c}{-}             & \multicolumn{1}{c}{29}                  & \multicolumn{1}{c}{5}              & \multicolumn{1}{c}{0.6}                    & \multicolumn{1}{c}{0.01}                   & \multicolumn{1}{c}{32}                  & 34                   \\ 
Combined                          & \multicolumn{1}{c}{-} & \multicolumn{1}{c}{42}                  & \multicolumn{1}{c}{13}             & \multicolumn{1}{c}{0.7}                    & \multicolumn{1}{c}{0.01}                   & \multicolumn{1}{c}{256}                  & 40                   \\ \hline

\hline
\end{tabular}
}
\end{table}

\subsection{Classification models}
\label{sec:trainedclassifiers}

\subsubsection{Deep neural decision tree}
The deep neural decision tree (DNDT) is a classifier structured as a tree \citep{7410529}, encompassing both decision and prediction nodes. Decision nodes, positioned within the tree but not at its leaves, serve as points where the tree assesses data features or conditions to make determinations. The decisions at each node guide the path a sample takes through the tree. On the other hand, prediction nodes are the leaf nodes of the tree, where the final prediction is generated. These nodes serve as the terminal points for predictions. Each prediction node corresponds to a specific class or outcome the classifier aims to predict. To classify a sample, the DNDT guides it to a leaf node, using a probability distribution to make the final prediction. The final prediction for a sample is determined by Equation \ref{equation6},

\begin{equation} \label{equation6}
 \mathbb{P}_{T}[y|x,\theta,\pi] = \sum\limits_{l \in L}^{} \pi_{ly} \mu_{l} (x|\theta)
\end{equation}
Where $\pi_{ly}$ is the likelihood of a sample arriving at leaf node $l$ to get placed in class $y$ and routing function $\mu_{l} (x|\theta)$ is the likelihood of a sample $x$ will arrive at leaf node $l$. 

To explicitly define the routing function, we introduce two binary relations determined by the tree's structure: \( \ell \swarrow n \), which holds true if \( \ell \) is part of the left subtree of node \( n \), and \( n \searrow \ell \), which holds true if \( \ell \) is part of the right subtree of node \( n \). Using these relations, \( \mu_\ell \) is expressed as shown in Equation \ref{equation7}. 


\begin{equation} \label{equation7}
\mu_\ell(x|\Theta) = \prod_{n \in \mathbb{N}} 
d_n(x; \Theta)^{\mathbf{1}_{\ell \swarrow n}} \overline{d_n}{(x; \Theta)}^{\mathbf{1}_{n \searrow \ell}}
\end{equation}
Where \( \mathbf{1}_P \) denotes an indicator function that activates when the condition \( P \) is satisfied, and \( d_n \) represents the decision output at the \( n \)-th node, calculated as shown in Equation \ref{equation82}, indicating a split or routing decision.

\begin{equation} \label{equation82}
d_n(x; \Theta) = \sigma(f_n(x; \Theta))
\end{equation}

The sigmoid function \( \sigma \) outputs a value between 0 and 1, used for binary decisions, while \( f_n \) is a learned function at the \( n \)-th node, often derived from a fully connected layer, that processes input \( x \) to determine the split decision. Additionally, \( \overline{d_n}(x; \Theta) \) is the complement of \( d_n(x; \Theta) \), calculated as \( 1 - d_n(x; \Theta) \), representing the alternative path or decision when the initial \( d_n \) decision is not chosen.


\subsubsection{Deep neural decision forest}
The deep neural decision forest (DNDF) \citep{7410529} is a classifier consisting of multiple DNDTs trained simultaneously. The DNDF produces its final output by averaging the individual outputs from each of the trees within the forest. The output of the DNDF is represented by Equation \ref{equation8}.

\begin{equation} \label{equation8}
 \mathbb{P}_{\emph{F}}[y|x] = \dfrac{1}{k} \sum\limits_{h=1}^{k} \mathbb{P}_{Th} [y|x]
\end{equation}
where $k$ represents how many trees are in the forest, $\mathbb{P}_{\emph{F}}[y|x]$ is the output for sample $x$ produced by the DNDF, and $\mathbb{P}_{Th} [y|x]$ is the output for sample $x$ produced by the DNDT. 

\subsection{Threshold moving}
In binary classification, the threshold used to determine class labels based on predicted probabilities is crucial for the model's performance. To improve class label assignment, we use the threshold moving technique, as relying on the default threshold of 0.50 often leads to suboptimal results. With this method, the ROC-AUC score is considered while determining the optimal threshold for a binary classifier. In medicine, the ROC-AUC score is a commonly used assessment metric, particularly for evaluating the effectiveness of diagnostic procedures \citep{article_tm}. ROC-AUC score-based threshold optimization is achieved by cross-validation tests. We calculate ROC-AUC scores over a threshold value range of 0.1 to 1, using 0.001 increments. The best threshold is then determined by selecting the one that produced the highest ROC-AUC score.

\subsection{Training strategies}
\label{sec:trainingstrategies}

We introduce five strategies to evaluate the effectiveness of different components of our proposed method: strategy 1, strategy 2, strategy 3, strategy 4, and strategy 5. Table \ref{tab:table2} shows the combinations of different training strategies used in each strategy. Strategy 1 exclusively relies on our trained classifiers. In strategy 2, we only use the threshold moving technique (to select the optimal threshold based on the ROC-AUC score) with a trained classifier. In strategy 3, we use both the threshold moving technique and the feature dimension reduction technique (to determine the key features using the RFECV method and Extra-Trees classifier) with a trained classifier. In strategy 4, we use the threshold moving technique, optimal feature selection using the RFECV method and the Extra-Trees classifier, and Bayesian Optimization (to select the best hyper-parameters). In strategy 5, we use the threshold moving technique, optimize feature selection using the RFECV method, use Bayesian Optimization, and apply SMOTE to balance the data of the minority class in an imbalanced dataset.

\begin{table}[!ht]
\centering
\caption{Combinations of several training strategies.
}
\label{tab:table2}
\begin{tabular}{lcccccc}
\hline
Strategy  &   \makecell[c]{Feature selection \\ (RFECV)} & \makecell[c]{Hyper-parameters\\selection method} & \makecell[c]{Up-sampling\\ (SMOTE)} & \makecell[c]{Threshold\\ moving} \\ \hline \hline
Strategy 1                                                                                                                                                                                                     & \ding{55}                                            & Default                                                                                  & \ding{55} & \ding{55}                                                   \\ 
Strategy 2                                                                                                                                                                                 & \ding{55}                                            & Default                                                                                  & \ding{55} & \checkmark                                                   \\ 
Strategy 3                                                                                                                                                                  & \checkmark                                            & Default                                                                                  & \ding{55}  & \checkmark                                                  \\ 
Strategy 4                                                                                                                                                  & \checkmark                                            & Bayesian Optimization                                                                 & \ding{55}      & \checkmark                                              \\ 
Strategy 5                                                                                                                                                             & \checkmark                       & Bayesian Optimization                                            & \checkmark        & \checkmark                                             \\ \hline

\hline
\end{tabular}
\end{table}

All strategies, except strategy 1, use the threshold moving technique to determine the optimal threshold based on ROC-AUC score. We also use 10-fold stratified cross-validation to assess the performance of our trained classifiers in all strategies. However, benchmark datasets conspicuously indicate a limited representation of the positive COVID-19 category, potentially harming the ML classifier's performance. To enhance the classifier's performance, we have integrated SMOTE during training in strategy 5 to balance the dataset. The difference among strategies 1, 2, 3, 4, and 5 is that strategies 1 and 2 do not use the feature selection method (RFECV) during the training phase, while strategies 3, 4, and 5 do. In addition, strategies 1, 2, and 3 use default hyper-parameters during the training phase, while strategies 4 and 5 use Bayesian Optimization to select the best hyper-parameters. Moreover, in strategy 5, SMOTE is the sole technique used for up-sampling, while the other strategies do not incorporate any up-sampling methods.

\section{Results and discussion}
\label{sec:results}

In this section, we present the results of our experiments on identifying COVID-19 by analyzing cough sounds. We first outline the evaluation metrics used to assess the performance of our proposed methods. Next, we present the classification performance of four strategies, which helps us to select the best strategy based on performance. Then, we describe the optimal feature selection techniques for selecting the most important features. Finally, we evaluate the effectiveness of our proposed method using state-of-the-art methods on different datasets.

\subsection{Evaluation metrics}
In our experimental evaluation, we use 10-fold stratified cross-validation to evaluate the performance using six standard evaluation metrics: ROC-AUC, Accuracy (Acc.), Precision, Recall/Sensitivity, Specificity (Spec.), and F1 score. The definitions of Precision, Recall/Sensitivity, Accuracy, Specificity, and F1 score are provided below:

\begin{center} 
$Accuracy = \dfrac{TP+TN}{TP+FP+FN+TN}$
\end{center}
\begin{center} 
$Precision = \dfrac{TP}{TP+FP}$
\end{center}
\begin{center} 
$Recall/Sensitivity = \dfrac{TP}{TP+FN}$
\end{center}
\begin{center} 
$Specificity = \dfrac{TN}{TN+FP}$
\end{center}
\begin{center} 
$F1\hspace{0.1cm}Score = 2\times\dfrac{Precision \times Recall}{Precision+Recall}$
\end{center}

Here, True Positive, False Positive, True Negative and False Negative are each represented as TP, FP, TN and FN, respectively. AUC, which stands for \quotes{Area Under the Curve}, signifies the area under the ROC curve, a probability curve. ROC, or the \quotes{Receiver Operating Characteristic}, is determined by the TPR (true positive rate) as a fraction of the FPR (false positive rate).

\begin{center} 
$TPR = \dfrac{TP}{TP+FN}$
\end{center}
\begin{center} 
$FPR = \dfrac{FP}{FP+TN}$
\end{center}

\subsection{Classification performance of five strategies}
\label{Classificationperformanceoffourstrategies}

All five strategies use DNDT and DNDF classifiers, and strategy 2 through strategy 5 incorporate the threshold moving technique. Strategies 3, 4, and 5 implement the RFECV method, using the Extra-Trees classifier to identify the most relevant features. Strategies 4 and 5 further integrate Bayesian Optimization to fine-tune Hyper-parameters for the trained classifiers. Additionally, in strategy 5, SMOTE is applied during the training phase to address data imbalance by generating synthetic samples for both positive and negative instances of cough related to COVID-19. The AUC-based classification performance of all five strategies is outlined in Table \ref{tab: comparison_among_all_strategies}.
\begin{table}[!ht]
\centering
\caption{Comparison of all strategies.}
\label{tab: comparison_among_all_strategies}
\resizebox{\textwidth}{!}{
\begin{tabular}{lccccccc}
\hline
\multirow{2}{*}{Dataset}  & \multirow{2}{*}{Category} & \multirow{2}{*}{\makecell[c]{Classification\\method}} & \multicolumn{5}{c}{AUC}                                                                                                                                                               \\ \cline{4-8} 
                                   &                                    &                                                                                           & Strategy 1 & Strategy 2 & Strategy 3 & Strategy 4 & Strategy 5 \\ \hline\hline 
\multirow{4}{*}{Cambridge}         & \multirow{2}{*}{Asymptomatic}      & DNDT                                                                                      & \multicolumn{1}{c}{0.55}                    & \multicolumn{1}{c}{0.72}                & \multicolumn{1}{c}{0.72}                & \multicolumn{1}{c}{0.92}                & 0.95                \\ 
                                   &                                    & DNDF                                                                                      & \multicolumn{1}{c}{0.51}                    & \multicolumn{1}{c}{0.72}                & \multicolumn{1}{c}{0.70}                & \multicolumn{1}{c}{0.96}                & 0.97                \\ \cline{2-8} 
                                   & \multirow{2}{*}{Symptomatic}       & DNDT                                                                                      & \multicolumn{1}{c}{0.77}                    & \multicolumn{1}{c}{0.94}                & \multicolumn{1}{c}{0.95}                & \multicolumn{1}{c}{0.95}                & 0.97                \\ 
                                   &                                    & DNDF                                                                                      & \multicolumn{1}{c}{0.69}                    & \multicolumn{1}{c}{0.93}                & \multicolumn{1}{c}{0.93}                & \multicolumn{1}{c}{0.97}                & 0.98                \\ \hline
\multirow{2}{*}{Coswara}           & \multirow{2}{*}{-}                 & DNDT                                                                                      & \multicolumn{1}{c}{0.51}                    & \multicolumn{1}{c}{0.62}                & \multicolumn{1}{c}{0.63}                & \multicolumn{1}{c}{0.79}                & 0.84                \\  
                                   &                                    & DNDF                                                                                      & \multicolumn{1}{c}{0.50}                    & \multicolumn{1}{c}{0.65}                & \multicolumn{1}{c}{0.62}                & \multicolumn{1}{c}{0.92}                & 0.92                \\ \hline
\multirow{2}{*}{COUGHVID}          & \multirow{2}{*}{-}                 & DNDT                                                                                      & \multicolumn{1}{c}{0.65}                    & \multicolumn{1}{c}{0.72}                & \multicolumn{1}{c}{0.72}                & \multicolumn{1}{c}{0.80}                & 0.81                \\  
                                   &                                    & DNDF                                                                                      & \multicolumn{1}{c}{0.63}                    & \multicolumn{1}{c}{0.70}                & \multicolumn{1}{c}{0.70}                & \multicolumn{1}{c}{0.92}                & 0.93                \\ \hline
\multirow{2}{*}{Virufy}            & \multirow{2}{*}{-}                 & DNDT                                                                                      & \multicolumn{1}{c}{0.85}                    & \multicolumn{1}{c}{0.94}                & \multicolumn{1}{c}{0.89}                & \multicolumn{1}{c}{0.95}                & 0.98                \\  
                                   &                                    & DNDF                                                                                      & \multicolumn{1}{c}{0.82}                    & \multicolumn{1}{c}{0.93}                & \multicolumn{1}{c}{0.88}                & \multicolumn{1}{c}{0.99}                & 0.99                \\ \hline
\multirow{2}{*}{Virufy + NoCoCoDa} & \multirow{2}{*}{-}                 & DNDT                                                                                      & \multicolumn{1}{c}{0.86}                    & \multicolumn{1}{c}{0.93}                & \multicolumn{1}{c}{0.93}                & \multicolumn{1}{c}{0.99}                & 0.99                \\  
                                   &                                    & DNDF                                                                                      & \multicolumn{1}{c}{0.79}                    & \multicolumn{1}{c}{0.93}                & \multicolumn{1}{c}{0.93}                & \multicolumn{1}{c}{0.99}                & 0.99                \\ \hline

                                   \hline
\end{tabular}
}
\end{table}

Across all datasets, strategy 2 consistently outperforms strategy 1 in terms of AUC, underscoring the effectiveness of the threshold moving technique. Specifically, when using the DNDT classifier, strategy 2 achieves superior AUC scores of 0.72, 0.94, 0.62, 0.72, 0.94, and 0.93 for the Cambridge asymptomatic, Cambridge symptomatic, Coswara, COUGHVID, Virufy, and Virufy merged with NoCoCoDa datasets, respectively. Similarly, with the DNDF classifier, strategy 2 surpasses strategy 1 with higher AUC scores of 0.72, 0.93, 0.65, 0.70, 0.93, and 0.93 for the same datasets.

From various datasets, strategy 3 demonstrates superior overall performance compared to strategy 1 and strategy 2. In particular, in the Cambridge symptomatic dataset, strategy 3, using the DNDT classifier, outshines strategy 1 and strategy 2, achieving a higher AUC score of 0.95. Similarly, in the Coswara dataset, strategy 3, using the DNDT classifier, outperforms the other two strategies with a higher AUC of 0.63.

The comparative analysis of Strategies 1, 2, 3, and 4, shows that strategy 4 consistently excels over the other three across almost all datasets. Specifically, using the DNDT classifier, strategy 4 achieves higher AUC scores of 0.92, 0.79, 0.80, 0.95, and 0.99 for the Cambridge asymptomatic, Coswara, COUGHVID, Virufy, and Virufy merged with the NoCoCoDa datasets, respectively. Furthermore, when using the DNDF classifier, strategy 4 consistently shows superior performance compared to the other strategies across all datasets, achieving AUC values of 0.96, 0.97, 0.92, 0.92, 0.99, and 0.99 for the Cambridge asymptomatic, Cambridge symptomatic, Coswara, COUGHVID, Virufy, and Virufy merged with the NoCoCoDa datasets, respectively.

When evaluating Strategies 1, 2, 3, 4, and 5, it is apparent that strategy 5 consistently outperforms the other four across nearly all datasets, demonstrating its effectiveness in COVID-19 classification. In the Cambridge asymptomatic, Cambridge symptomatic, and COUGHVID datasets, strategy 5, which uses both DNDT and DNDF classifiers, outperforms Strategies 1, 2, 3, and 4 by achieving higher AUC scores with both classifiers. Specifically, strategy 5 attains AUC scores of 0.95 and 0.97 for the Cambridge asymptomatic, 0.97 and 0.98 for the Cambridge symptomatic, 0.84 and 0.92 for the Coswara, 0.81 and 0.93 for the COUGHVID, 0.98 and 0.99 for the Virufy, and 0.99 and 0.99 for the Virufy + NoCoCoDa dataset with the DNDT and DNDF classifiers, respectively. 
These results clearly demonstrate the superior performance of Strategy 5 in comparison to the other strategies. Notably, the DNDF classifier consistently surpasses the DNDT classifier in all datasets, further highlighting the enhanced performance of Strategy 5. This comprehensive outperformance by strategy 5, driven by the use of both classifiers and advanced techniques such as SMOTE, Bayesian Optimization, and RFECV, solidifies its position as the most robust method for categorizing COVID-19 from cough sounds. A detailed breakdown of strategy 5’s classification performance across all evaluation metrics and datasets is provided in Table \ref{tab:strategy5}.
\begin{table}[!ht]
\centering
\caption{Classification performance of strategy 5 (RFECV+BO+SMOTE+TM).}
\label{tab:strategy5}
\begin{tabular}{lclcccccc}
\hline
Dataset                                   & Category           & Method                                                                                   & Accuracy & AUC  & Precision & Recall & F1 & Specificity \\ \hline \hline
\multirow{4}{*}{Cambridge}                                                     & \multirow{2}{*}{Asymptomatic} & \begin{tabular}[c]{@{}l@{}}DNDT\end{tabular} & 0.96 & 0.95 & 0.92      & 0.95            & 0.93                                               & 0.96        \\ 
                                                                               &                              & \begin{tabular}[c]{@{}l@{}}DNDF\end{tabular} & 0.98 & 0.97 & 1               & 0.94   & 0.96                                               & 1        \\ \cline{2-9} 
                                                                               & \multirow{2}{*}{Symptomatic}  & \begin{tabular}[c]{@{}l@{}}DNDT\end{tabular} & 0.96 & 0.97 & 1               & 0.93   & 0.96                                               & 1                 \\ 
                                                                               &                              & \begin{tabular}[c]{@{}l@{}}DNDF\end{tabular} & 0.98          & 0.98 & 1         & 0.97            & 0.98                                                        & 1           \\ \hline
\multirow{2}{*}{Coswara}                                                       & \multirow{2}{*}{-}           & \begin{tabular}[c]{@{}l@{}}DNDT\end{tabular} & 0.86          & 0.84          & 0.52               & 0.80   & 0.62                                                        & 0.87                 \\  
                                                                               &                              & \begin{tabular}[c]{@{}l@{}}DNDF\end{tabular} & 0.92 & 0.92 & 0.72      & 0.93            & 0.80                                               & 0.91      \\ \hline
\multirow{2}{*}{COUGHVID}                                                      & \multirow{2}{*}{-}           & \begin{tabular}[c]{@{}l@{}}DNDT\end{tabular} & 0.81          & 0.81          & 0.83               & 0.79            & 0.78                                                        & 0.83                 \\ 
                                                                               &                              & \begin{tabular}[c]{@{}l@{}}DNDF\end{tabular} & 0.93 & 0.93 & 0.93      & 0.94   & 0.93                                               & 0.93        \\ \hline
\multirow{2}{*}{Virufy}                                                        & \multirow{2}{*}{-}           & \begin{tabular}[c]{@{}l@{}}DNDT\end{tabular} & 0.98 & 0.98 & 1               & 0.96      & 0.98                                               & 1                 \\  
                                                                               &                              & \begin{tabular}[c]{@{}l@{}}DNDF\end{tabular} & 0.99          & 0.99          & 0.98         & 1            & 0.99                                                        & 0.99           \\ \hline
\multirow{2}{*}{\begin{tabular}[c]{@{}l@{}}Virufy + NoCoCoDa\end{tabular}} & \multirow{2}{*}{-}           & \begin{tabular}[c]{@{}l@{}}DNDT\end{tabular} & 0.99          & 0.99          & 1         & 0.99            & 0.99                                                        & 1          \\  
                                                                               &                              & \begin{tabular}[c]{@{}l@{}}DNDF\end{tabular} & 0.99 & 0.99 & 1         & 0.99   & 0.99                                             & 1           \\ \hline

                                                                               \hline

\end{tabular}
\end{table}

\begin{figure}[!ht]
     \centering
     \begin{subfigure}[b]{0.33\textwidth}
         \centering
         \includegraphics[width=\textwidth]{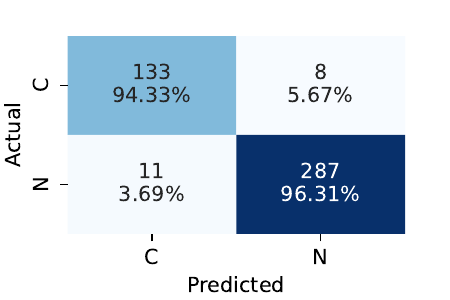}
         \caption{}
         \label{fig:a1}
     \end{subfigure}
     \hfill
     \begin{subfigure}[b]{0.33\textwidth}
         \centering
         \includegraphics[width=\textwidth]{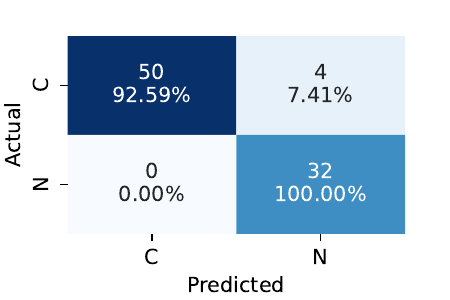}
         \caption{}
         \label{fig:b1}
     \end{subfigure}
     \hfill
     \begin{subfigure}[b]{0.33\textwidth}
         \centering
         \includegraphics[width=\textwidth]{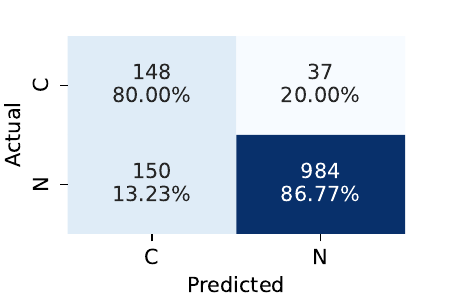}
         \caption{}
         \label{fig:c1}
     \end{subfigure}
     
     \begin{subfigure}[b]{0.33\textwidth}
         \centering
         \includegraphics[width=\textwidth]{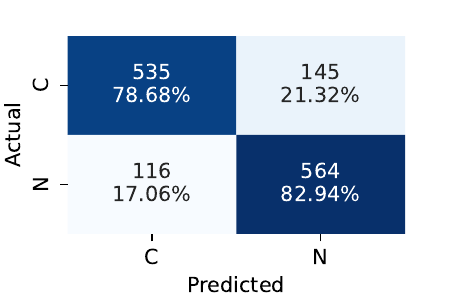}
         \caption{}
         \label{fig:d1}
     \end{subfigure}
     \hfill
     \begin{subfigure}[b]{0.33\textwidth}
         \centering
         \includegraphics[width=\textwidth]{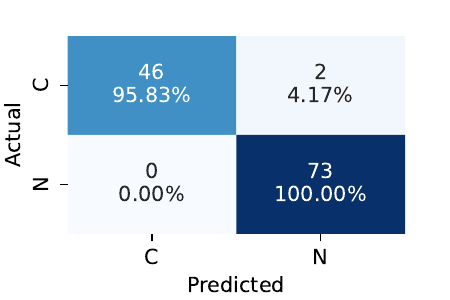}
         \caption{}
         \label{fig:e1}
     \end{subfigure}
     \hfill
     \begin{subfigure}[b]{0.33\textwidth}
         \centering
         \includegraphics[width=\textwidth]{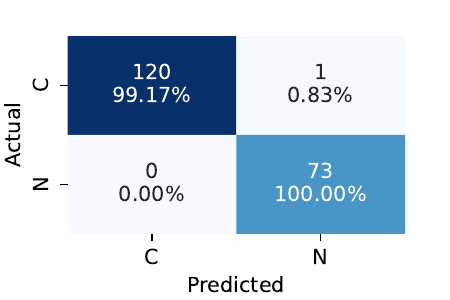}
         \caption{}
         \label{fig:f1}
     \end{subfigure}
     \caption{The confusion matrices for strategy 5, using the Deep Neural Decision Tree (DNDT) classifier with 10-fold cross-validation: (a) Cambridge asymptomatic, (b) Cambridge symptomatic, (c) Coswara, (d) COUGHVID, (e) Virufy, and (f) Virufy merged with NoCoCoDa. ``C" represents COVID-19 Positive and ``N" represents Non COVID-19 cough instances.}
     \label{fig:confusionmatrixsDNDT}
\end{figure}

\begin{figure*}[!ht]
     \centering
     \begin{subfigure}[b]{0.33\textwidth}
         \centering
         \includegraphics[width=\textwidth]{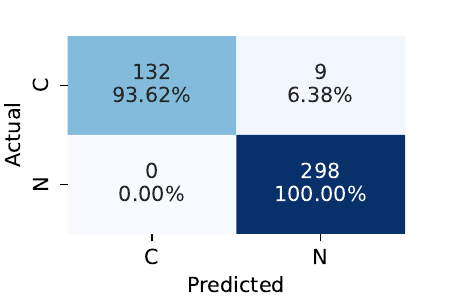}
         \caption{}
         \label{fig:a2}
     \end{subfigure}
     \hfill
     \begin{subfigure}[b]{0.33\textwidth}
         \centering
         \includegraphics[width=\textwidth]{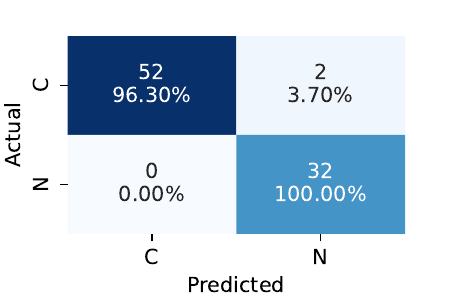}
         \caption{}
         \label{fig:b2}
     \end{subfigure}
     \hfill
     \begin{subfigure}[b]{0.33\textwidth}
         \centering
         \includegraphics[width=\textwidth]{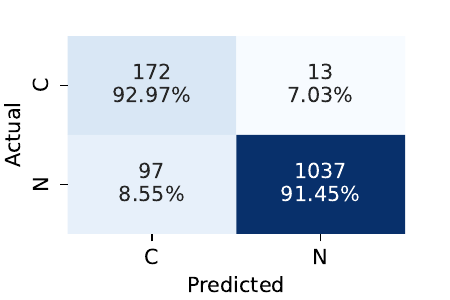}
         \caption{}
         \label{fig:c2}
     \end{subfigure}
     
     \begin{subfigure}[b]{0.33\textwidth}
         \centering
         \includegraphics[width=\textwidth]{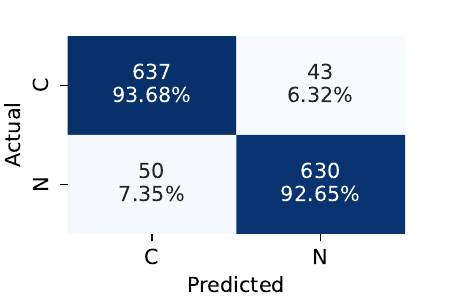}
         \caption{}
         \label{fig:d2}
     \end{subfigure}
     \hfill
     \begin{subfigure}[b]{0.33\textwidth}
         \centering
         \includegraphics[width=\textwidth]{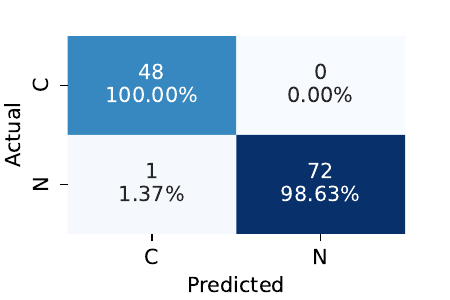}
         \caption{}
         \label{fig:e2}
     \end{subfigure}
     \hfill
     \begin{subfigure}[b]{0.33\textwidth}
         \centering
         \includegraphics[width=\textwidth]{confmatrix_VirufyNoCoCoDa_DNDTF_last.pdf}
         \caption{}
         \label{fig:f2}
     \end{subfigure}
 \caption{The confusion matrices for strategy 5, using the Deep Neural Decision Forest (DNDF) classifier with 10-fold cross-validation: (a) Cambridge asymptomatic, (b) Cambridge symptomatic, (c) Coswara, (d) COUGHVID, (e) Virufy, and (f) Virufy merged with NoCoCoDa datasets. ``C" represents COVID-19 Positive, and ``N" represents Non COVID-19 cough instances.}
 \label{fig:confusionmatrixsDNDF}
\end{figure*}

Figure \ref{fig:confusionmatrixsDNDT} illustrates the confusion matrices for the DNDT classifier, using strategy 5 to evaluate its effectiveness in detecting COVID-19 across multiple datasets. Similarly, Figure \ref{fig:confusionmatrixsDNDF} presents the confusion matrices for the DNDF classifier with strategy 5, detailing its performance in COVID-19 detection on the same datasets.

For the Cambridge asymptomatic dataset, the DNDT classifier correctly identifies 133 out of 141 COVID-19 positive cough samples and 287 out of 298 cough samples from healthy individuals, as depicted in Figure~\ref{fig:a1}. In comparison, the DNDF classifier achieves a slightly improved performance by accurately predicting 132 out of 141 COVID-19 positive cough samples and all 298 cough samples from healthy individuals, as shown in Figure~\ref{fig:a2}.

In the Cambridge symptomatic dataset, the DNDT classifier demonstrates robust performance by correctly classifying all 32 cough samples from healthy individuals and 50 out of 54 COVID-19 positive cough samples (Figure~\ref{fig:b1}). The DNDF classifier achieves comparable results, successfully predicting 96.30\% (52 out of 54) of COVID-19 positive cough samples and 100\% of healthy cough samples (Figure~\ref{fig:b2}).

For the Coswara dataset, the DNDT classifier demonstrates strong predictive performance, correctly classifying 984 out of 1,134 cough samples from healthy individuals (86.81\%) and 148 out of 185 COVID-19 positive cough samples (80\%) (Figure~\ref{fig:c1}). In comparison, the DNDF classifier achieves higher accuracy, correctly predicting 91.45\% (1,037 out of 1,134) of cough samples from healthy individuals and 92.97\% (172 out of 185) of COVID-19 positive cough samples (Figure~\ref{fig:c2}).

In the COUGHVID dataset, the DNDT classifier correctly identifies 535 out of 680 COVID-19 positive cough samples (78.68\%) and 564 out of 680 healthy cough samples (82.94\%) (Figure~\ref{fig:d1}). The DNDF classifier demonstrates improved accuracy, correctly predicting 637 out of 680 COVID-19 positive cough samples (93.68\%) and 630 out of 680 healthy cough samples (92.65

For the Virufy dataset, the DNDT classifier achieves high accuracy, correctly classifying 46 out of 48 COVID-19 positive cough samples (95.83\%) and all 73 healthy cough samples (100\%) (Figure~\ref{fig:e1}). The DNDF classifier shows comparable results, accurately identifying 72 out of 73 healthy cough samples (98.63\%) and all 48 COVID-19 positive cough samples (100\%) (Figure~\ref{fig:e2}).

When the Virufy dataset is integrated with the NoCoCoDa dataset, both the DNDT and DNDF classifiers exhibit exceptional accuracy, correctly predicting 120 out of 121 COVID-19 positive cough samples (99.17\%) and all 73 healthy cough samples (100\%) (Figures~\ref{fig:f1} and~\ref{fig:f2}).

The comparative analysis of DNDT and DNDF classifiers across multiple datasets reveals a consistent pattern of DNDF outperforming DNDT in terms of predictive accuracy, particularly for COVID-19 positive cough samples. While DNDT demonstrates robust performance with high accuracy across most datasets, DNDF achieves superior results by consistently providing higher classification rates for both COVID-19 positive and healthy samples. For instance, on the Coswara dataset, the DNDF classifier achieves higher classification rates for both healthy and COVID-19 positive cough samples compared to DNDT, underscoring its greater sensitivity and specificity. Similarly, on the COUGHVID dataset, the DNDF classifier significantly improves COVID-19 positive sample detection accuracy, achieving 93.68\% compared to the DNDT’s 78.68\%, a difference that could be critical in real-world applications where false negatives carry severe consequences. Furthermore, the DNDF classifier exhibits greater consistency across smaller datasets, such as Virufy, where it attains near-perfect accuracy, further emphasizing its robustness. The DNDF classifier's consistent superiority suggests that the decision forest architecture is better suited for capturing the complex patterns in cough sound data, making it a more reliable tool for COVID-19 detection. 
This enhanced performance of DNDF can be attributed to its integration of decision forest structures, which leverage the ensemble approach to reduce prediction variance and improve generalization. In contrast, DNDT relies on a single decision tree model, which, while effective, is inherently more susceptible to overfitting and less adaptable to complex variations in the data. These findings suggest that DNDF's ensemble-based architecture offers a more robust and reliable framework for COVID-19 detection from cough sounds.

\subsection{Comparative analysis of performance with state-of-the-art methods}
\label{comparison}

Table \ref{tab:Table 7} provides a comparative analysis of our innovative approaches for COVID-19 diagnosis from cough samples in contrast to contemporary methods. Our methodologies encompass several pivotal components, including feature selection via Recursive Feature Elimination with Cross-Validation (RFECV), Hyper-parameter tuning through Bayesian Optimization (BO), data augmentation using the Synthetic Minority Over-sampling Technique (SMOTE) during training, and dynamic threshold adjustment guided by ROC-AUC score (TM). It is worth noting that the comparative assessment is conducted with a focus on three specific prior studies, namely, \citet{10.1145/3394486.3412865}, \citet{article6} and \citet{article2}. The rationale behind this limitation is rooted in the availability of detailed implementation specifics and datasets for these select works, which ensures a more accurate and comprehensive comparative analysis. In this comparison, we exclude studies that utilize merged datasets for training or testing instead of separate datasets like ours. Additionally, studies employing multi-class classification, significantly different datasets, or omitting essential metrics such as AUC, Precision, Recall, or F1-score are excluded. For the Cambridge dataset, we focus on two distinct categories: Asymptomatic and Symptomatic. Studies that fail to adopt these specific categories or use alternative dataset categorizations are also excluded from the comparison. 
\begin{table}[!htp]
\centering
\footnotesize
\caption{A comparison between our proposed method and state-of-the-art methods.}
\label{tab:Table 7}
\begin{tabular}{lclccc}
\hline
Dataset                                                                 & Category            & Method                                                                                   & AUC  & Precision & Recall \\ \hline\hline
\multirow{13}{*}{Cambridge}                                                      & \multirow{6}{*}{Asymptomatic} & \citet{10.1145/3394486.3412865}                                                                                      & 0.80           & 0.72               & 0.69            \\ 
 &                              &  \citet{aytekin2023covid19}                                                                   & \textbf{0.97}          & 0.92               & 0.94             \\
                                                                                 &                              & 
                                                                                 \citet{article2}                                                                 & 0.88 & 0.75               & 0.81            \\
&                              &  \citet{dentamaro2022auco}                                                                                   & 0.83             & 0.80               & 0.80            \\

                                                                                 &                              & Proposed (DNDT+RFECV+BO+SMOTE+TM) & 0.95 & 0.92      & \textbf{0.95}   \\  
                                                                                 &                              & Proposed (DNDF+RFECV+BO+SMOTE+TM) & \textbf{0.97}         & \textbf{1}      & 0.94   \\ \cline{2-6} 
                                                                                 & \multirow{7}{*}{Symptomatic}  &   \citet{10.1145/3394486.3412865}                                                                                      & 0.87          & 0.70                & 0.90             \\ 
                                                                                 &                              &  \citet{chowdhury2021qucoughscope}                                                                                   & -             & 0.87               & 0.82            \\ 
                                                                                   &                              &  \citet{aytekin2023covid19}                                                                   & \textbf{0.98}          & 0.94               & 0.93             \\  
                                                                                
                                                                                 &                              &   \citet{article2}                                                                & 0.95          & \textbf{1}         & 0.91            \\ 
                                                                                  &                              &  \citet{dentamaro2022auco}                                                                                   & 0.93             & 0.89               & 0.93            
                                                                                  \\

                                                                                 &                              & Proposed (DNDT+RFECV+BO+SMOTE+TM) & 0.97 & \textbf{1}               & 0.93      \\ 
                                                                                 &                              & Proposed (DNDF+RFECV+BO+SMOTE+TM) & \textbf{0.98}          & \textbf{1}               & \textbf{0.97}           \\ \hline
\multirow{5}{*}{Coswara}                                                         & \multirow{5}{*}{-}           & \citet{zhang2022robust}                                                                & 0.86          & -                & 0.60            \\ 
&  & \citet{sobahi2022explainable}                                                                & -          & \textbf{0.90}                & 0.88            \\ 
&  & \citet{article2}                                                                & 0.66          & 0.76                & 0.47            \\ 
                                                                                 
                                                                                 &                              & Proposed (DNDT+RFECV+BO+SMOTE+TM) & 0.84          & 0.52               & 0.80  \\  
                                                                                 &                              & Proposed (DNDF+RFECV+BO+SMOTE+TM) & \textbf{0.92} & 0.72      & \textbf{0.93}            \\ \hline
\multirow{8}{*}{COUGHVID}                                                        & \multirow{8}{*}{-}  
& \citet{Orlandic2022ASA} 
                 & 0.88 &    -   & -            \\
& & \citet{Sunitha2022ACA} 
                 & - &    0.76   & 0.77            \\
& & \citet{hamdi2022attention} 
                 & 0.91 &    0.91   & 0.90            \\
& & \citet{inbook} 
                 & 0.91 &    \textbf{0.95}   & 0.86            \\
 &                              &  \citet{aytekin2023covid19}                                                                   & 0.83          & 0.77               & \textbf{1}             \\                
& & \citet{s23114996} 
                 & 0.76 &    0.69   & 0.68            \\
                 
                 &
                 & Proposed (DNDT+RFECV+BO+SMOTE+TM) & 0.81          & 0.83                & 0.79             \\ 
                                                                                 &                              & Proposed (DNDF+RFECV+BO+SMOTE+TM) & \textbf{0.93} & 0.93      & 0.94   \\ \hline
\multirow{10}{*}{Virufy}                                                          & \multirow{10}{*}{-}           & \citet{article6} 
                 & - & \textbf{1}      & 0.95            \\
                 &
                 & \citet{kapoor2022cough} 
                 & - & 0.91      & 0.90            \\
                 &
                 & \citet{nafiz2023automated} 
                 & 0.93 & 0.86      & 0.80            \\
                 &
                 & \citet{ISLAM2022100025} 
                 & - & \textbf{1}      & 0.95            \\
                 &
                 & \citet{MelekManshouri2021IdentifyingCB} 
                 & - & 0.99      & -            \\
                 &
                 & \citet{erdougan2021covid} 
                 & - & 0.97      & 0.99            \\

                                                                                 &                              & \citet{sobahi2022explainable} & - & 0.99      & 0.97            \\
                                                                                 &                              &  \citet{article2}                                                                    & 0.94 & 0.89               & 0.98   \\
                                                                                 
                                                                                 &                              & Proposed (DNDT+RFECV+BO+SMOTE+TM) & 0.98 & \textbf{1}      & 0.96            \\ 
                                                                                 &                              & Proposed (DNDF+RFECV+BO+SMOTE+TM) & \textbf{0.99}          & 0.98              & \textbf{1}  \\ \hline
\multirow{4}{*}{\begin{tabular}[l]{@{}c@{}}Virufy+NoCoCoDa\end{tabular}} & \multirow{4}{*}{-}           & \citet{10.1007/s00521-021-06346-3}                                                                                               & \textbf{0.99} & 0.99               & 0.97            \\ 
                                                                                  
                                                                                 &                              &  \citet{article2}                                                                    & 0.98          & 0.99               & 0.98            \\ 
                                                                                 &                              & Proposed (DNDT+RFECV+BO+SMOTE+TM) & \textbf{0.99} & \textbf{1}               & \textbf{0.99}      \\  
                                                                                 &                              & Proposed (DNDF+RFECV+BO+SMOTE+TM) & \textbf{0.99}          & \textbf{1}               & \textbf{0.99}            \\ \hline

                                                                                 \hline
\multicolumn{6}{l}{- \textbf{Bold} values represent the highest scores.}  
\end{tabular}
\end{table}

Our proposed method demonstrates constant superiority over state-of-the-art techniques on a wide range of datasets, confirming its strong performance in COVID-19 diagnosis. On the Cambridge Asymptomatic dataset, \citet{aytekin2023covid19} and the DNDF classifiers achieve the highest AUC of 0.97. The DNDF classifier records a flawless precision score of 1, whereas the DNDT classifier is distinguished by the highest recall of 0.95. These results highlight the comprehensive and balanced performance of our proposed method on the Cambridge Asymptomatic dataset. In the Cambridge Symptomatic dataset, \citet{aytekin2023covid19}~and the DNDF classifiers both secure the highest AUC score at 0.98. The DNDF classifier records an impressive recall score of 0.97. Notable distinctions include the fact that \citet{article2}, along with both the DNDT and DNDF classifiers, attain the maximum precision score of 1, highlighting their accuracy in identifying COVID-19 instances.

The DNDF classifier performs exceptionally well on the Coswara dataset, achieving an impressive AUC score of 0.92 and a recall score of 0.93. The proposed method demonstrates relatively lower precision on the Coswara dataset compared to other datasets, primarily due to a significant data imbalance, with a much larger proportion of COVID-19 negative samples. Notably, with a score of 0.90, \citet{sobahi2022explainable} obtained the maximum precision on this dataset. On the COUGHVID dataset, the DNDF classifier surpasses other methods with an AUC of 0.93. Meanwhile, \citet{inbook} achieve a precision of 0.95, while \citet{aytekin2023covid19} reach a perfect recall score of 1. The DNDF classifier performs exceptionally well on the Virufy dataset, achieving outstanding AUC and recall scores of 0.99 and 1, respectively. In contrast, \citet{article6}, \citet{ISLAM2022100025} and the DNDT classifiers achieve a perfect precision score of 1. Moreover, in the Virufy dataset merged with the NoCoCoDa dataset, both the DNDT and DNDF classifiers attain a perfect precision and recall scores of 1 and 0.99, respectively. \citet{10.1007/s00521-021-06346-3}, along with both the DNDT and DNDF classifiers, achieve perfect AUC scores of 0.99. Compared to the DNDT classifier, the DNDF classifier performs better in terms of AUC and recall across the Cambridge Symptomatic, Coswara, COUGHVID, and Virufy datasets. However, on the Cambridge Asymptomatic and Virufy datasets, the DNDT classifier outperforms the DNDF classifier, especially in terms of recall and precision, respectively.

These commendable AUC, precision, and recall values collectively underscore the compelling effectiveness of our proposed method in the challenging task of classifying COVID-19 cases based on cough sound data. These results not only validate the robustness of our approach but also underscore its potential to make a significant contribution to advancing COVID-19 diagnostic capabilities.

\subsection{Cross-datasets evaluation}
\label{CrossSectionalStudy}

During our research, we conducted a comprehensive cross-datasets study to assess our proposed method’s effectiveness in diagnosing COVID-19 based on cough sounds. Our study encompassed a diverse range of COVID-19 cough datasets, which included the Cambridge (Asymptomatic), Cambridge (Symptomatic), Coswara, COUGHVID, Virufy, and Virufy with NoCoCoDa datasets. This comprehensive approach allows us to evaluate the robustness and generalizability of our method.
To ensure the integrity of our study, we followed a systematic approach. We trained our proposed method on one dataset, allowing it to learn from the unique characteristics of that dataset. Subsequently, we rigorously validated the model's performance on the remaining datasets, one dataset at a time. This validation process, performed independently for each dataset, is crucial in assessing the method's adaptability and its ability to distinguish COVID-19 cough sounds under varying conditions. The RFECV is intentionally taken out of our methodology, which is a noteworthy component. This was because different datasets have drastically different ideal feature properties that are necessary for a reliable COVID-19 diagnosis. We use BO to fine-tune hyper-parameters to improve the technique’s overall performance. In addition, we increased the data and addressed class imbalance, a frequent issue in medical datasets, by using SMOTE during the training phase. Deciding on the best classification threshold is made possible by the TM technique. We examine the DNDF classifier in this context.

The cross-datasets classification performance for our proposed method (DNDF+BO+SMOTE+TM) is reported in Table \ref{tab: tab11}. The Cambridge (Symptomatic) dataset performs better than other testing datasets when the Cambridge (Asymptomatic) dataset is used for training, obtaining AUC, precision, recall, and F1 scores of 0.89, 0.92, 0.91, and 0.92, respectively. Within the training dataset of Cambridge (Asymptomatic), remarkable performance metrics come to the forefront when assessing the performance of our method across various testing datasets. The testing datasets from Virufy + NoCoCoDa and Cambridge (Symptomatic) stand out in particular for their noteworthy accomplishments in terms of AUC, precision, recall, and F1 scores. In particular, the testing datasets for Virufy + NoCoCoDa and Cambridge (Symptomatic) yield AUC scores of 0.73 and 0.89, precision values of 0.80 and 0.92, and recall values of 0.79 and 0.91, respectively. Furthermore, the Virufy testing dataset provides excellent results, with an AUC of 0.70. Recall and F1 values of 0.73 and 0.65, respectively, support this outstanding performance and highlight the method's competence in differentiating COVID-19 cough sounds when trained on the Cambridge (Asymptomatic) dataset.
\begin{table}[!htp]
\centering
\caption{Classification performance of our proposed method (DNDF classifier with strategy 5) and the state-of-the-art methods in the cross-datasets study.}
\label{tab: tab11}
\begin{tabular}{llcccc}
\hline
Training Dataset               & Testing Dataset & AUC  & Precision & Recall & F1 Score \\ \hline \hline
\multirow{5}{*}{Cambridge (Asymptomatic)} & Cambridge (Symptomatic)     & 0.89          & 0.92               & 0.91            & 0.92              \\ & Coswara                  & 0.56          & 0.16               & 0.75            & 0.26              \\ & COUGHVID                 & 0.51          & 0.56               & 0.09            & 0.16              \\ & Virufy                   & 0.70 & 0.59               & 0.73   & 0.65     \\ & Virufy + NoCoCoDa        & 0.73          & 0.80      & 0.79            & 0.80              \\ \hline
\multirow{5}{*}{Cambridge (Symptomatic)}  & Cambridge (Asymptomatic)    & 0.69          & 0.50               & 0.72   & 0.59              \\ & Coswara                  & 0.56          & 0.17               & 0.60            & 0.27              \\ & COUGHVID                 & 0.52          & 0.51               & 0.80            & 0.62              \\ & Virufy                   & 0.69          & 0.61               & 0.65            & 0.63              \\ & Virufy + NoCoCoDa        & 0.72 & 0.90      & 0.55            & 0.68     \\ \hline
\multirow{7}{*}{Coswara}               & Cambridge (Asymptomatic)    & 0.53          & 0.34               & 0.87            & 0.48              \\ & Cambridge (Symptomatic)     & 0.52          & 0.80               & 0.07   & 0.14     \\ & COUGHVID                 & 0.53          & 0.57               & 0.24            & 0.33              \\ & Virufy                   & 0.58          & 0.48               & 0.60            & 0.53              \\ 
& Virufy + NoCoCoDa        & 0.59 & 0.72      & 0.49            & 0.58              \\ \hline
\multirow{5}{*}{COUGHVID}              & Cambridge (Asymptomatic)    & 0.54          & 0.34         & 0.91            & 0.50              \\ & Cambridge (Symptomatic)     & 0.55          & 0.65               & 0.98            & 0.79              \\ & Coswara                  & 0.57          & 0.17               & 0.85            & 0.28              \\ & Virufy                   & 0.60 & 0.54               & 0.44   & 0.48              \\ & Virufy + NoCoCoDa        & 0.55          & 0.67               & 0.59            & 0.63     \\ \hline
\multirow{5}{*}{Virufy}                & Cambridge (Asymptomatic)    & 0.61          & 0.39               & 0.82            & 0.53              \\ & Cambridge (Symptomatic)     & 0.67          & 0.75               & 0.74            & 0.75              \\ & Coswara                  & 0.55          & 0.16               & 0.76            & 0.26              \\ & COUGHVID                 & 0.53          & 0.59               & 0.20            & 0.30              \\ & Virufy + NoCoCoDa        & 0.85 & 0.90      & 0.86   & 0.88     \\ \hline
\multirow{5}{*}{Virufy + NoCoCoDa}     & Cambridge (Asymptomatic)    & 0.67          & 0.44               & 0.84            & 0.58              \\ & Cambridge (Symptomatic)     & 0.69          & 0.81      & 0.63            & 0.71              \\ & Coswara                  & 0.55          & 0.18               & 0.36   & 0.24              \\ & COUGHVID                 & 0.52          & 0.51               & 0.86            & 0.64              \\ & Virufy                   & 0.87 & 0.89               & 0.81            & 0.85     \\ \hline
                                                                         
\hline
\end{tabular}
\end{table}

Evaluating the Cambridge (Symptomatic) training dataset across diverse testing datasets, including the Virufy and Virufy + NoCoCoDa, reveals robust and noteworthy performance across multiple key metrics. Specifically, these datasets achieve impressive AUC, precision, recall, and F1 scores, highlighting their proficiency in distinguishing COVID-19 cough sounds. The testing datasets for Virufy and Virufy + NoCoCoDa perform well for the Cambridge (Symptomatic) training dataset. They obtain, respectively, precision values of 0.61 and 0.90, recall scores of 0.65 and 0.55, AUC scores of 0.69 and 0.72, and F1 scores of 0.63 and 0.68. The Cambridge (Asymptomatic) testing dataset also demonstrates commendable performance, particularly excelling in AUC with a score of 0.69, and achieving a notable recall score of 0.72. These results underscore the effectiveness and versatility of the method when applied to different datasets, further emphasizing its potential in COVID-19 cough sound classification.

The Coswara training dataset yields impressive outcomes when evaluated using the Virufy, and Virufy + NoCoCoDa testing datasets. These testing datasets showcase robust performance across essential metrics, including AUC, precision, recall, and F1 scores. Consequently, they attain noteworthy performance indicators with values of 0.58, 0.48, 0.60, and 0.53; and 0.59, 0.72, 0.49, and 0.58, respectively. The assessment of the COUGHVID training dataset demonstrates significant performance in various metrics, encompassing AUC, precision, recall, and F1 score. Notably, the Cambridge (Symptomatic) and Virufy + NoCoCoDa testing datasets stand out as robust achievers, securing scores of 0.55, 0.65, 0.98, and 0.79, and 0.55, 0.67, 0.59, and 0.63, respectively.

In the Virufy training dataset assessment, the Cambridge (Symptomatic) and Virufy + NoCoCoDa testing datasets perform well, securing AUC, precision, recall, and F1 scores of 0.67, 0.75, 0.74, and 0.75, and 0.85, 0.90, 0.86, and 0.88, respectively. Moreover, the Cambridge (Asymptomatic) testing dataset shows notable performance, particularly in AUC (0.61) and recall (0.82). The Virufy testing dataset performs well in the context of the Virufy integrated with NoCoCoDa training dataset, with scores of 0.87, 0.89, 0.81, and 0.85 for AUC, precision, recall, and F1 score, respectively. Furthermore, the Cambridge (Symptomatic) testing dataset shows commendable results, particularly in terms of AUC (0.69) and precision (0.81). The Cambridge (Asymptomatic) testing dataset performs well, especially in terms of AUC (0.67) and recall (0.84).

\begin{figure}[!ht]
\centering
\includegraphics[width=\textwidth, inner]{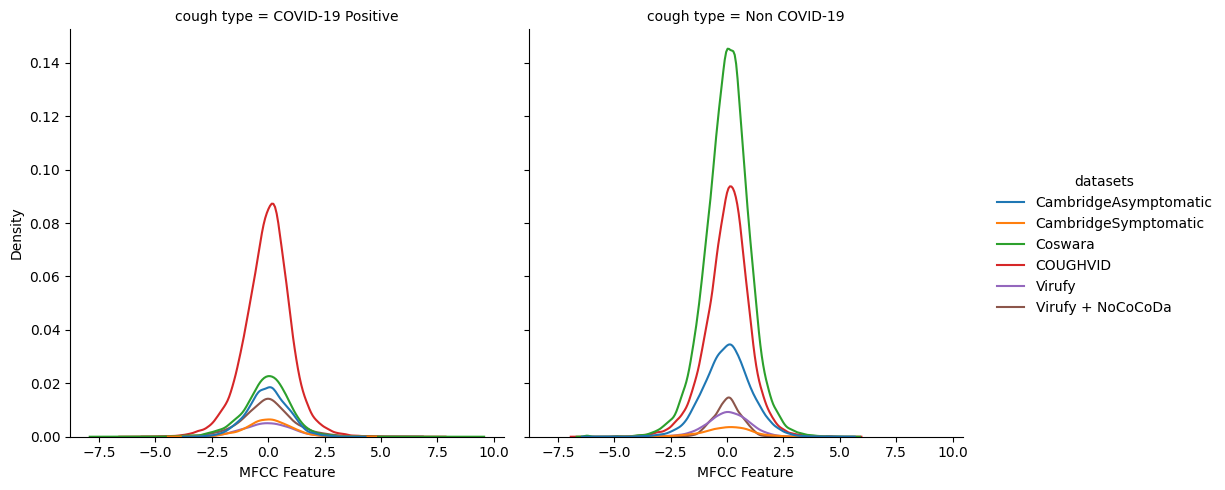}
\caption{The distribution of MFCC features across COVID-19 positive and negative samples across several datasets.}
\label{fig:mfcc}
\end{figure}

Using Kernel Density Estimation (KDE) curves, Figure \ref{fig:mfcc} illustrates the distribution of MFCC features across COVID-19 positive and negative samples across several datasets. The KDE curve's y-axis shows the estimated probability density at particular places along the x-axis, while the x-axis indicates the data points corresponding to MFCC features. The peak probability density in the distribution is represented by the highest point on the y-axis. In particular, peak probability densities of 0.088, 0.023, 0.019, 0.014, 0.006, and 0.004 are revealed by the KDE curve for MFCC features in COVID-19 positive cough samples for the COUGHVID, Coswara, Cambridge (Asymptomatic), Virufy merged with NoCoCoDa, Cambridge (Symptomatic), and Virufy datasets, respectively. Likewise, peak probability densities are also noted for non-COVID-19 cough samples, and they are 0.146, 0.094, 0.035, 0.014, 0.010, and 0.004 for the Coswara, COUGHVID, Cambridge (Asymptomatic), Virufy combined with NoCoCoDa, Virufy, and Cambridge (Symptomatic) datasets, respectively. The properties of the Cambridge (Asymptomatic), Cambridge (Symptomatic), Virufy, and Virufy + NoCoCoDa datasets are identical after examining the highest probable densities and shapes in Figure \ref{fig:mfcc}. In addition, it is noted that the training and testing datasets have similar features when assessing classification accuracy and looking at feature distribution.

As demonstrated in Table \ref{tab: tab11}, the cross-datasets evaluation reveals generally poor performance in classification tasks. However, there are notable exceptions: the evaluation shows promising results on the Virufy testing dataset when trained with the combined Virufy and NoCoCoDa dataset, on the combined Virufy and NoCoCoDa testing dataset when trained with the Virufy dataset, and on the Cambridge (Symptomatic) testing dataset when trained with the Cambridge (Asymptomatic) dataset during the cross-datasets study. The improved performance in the Virufy and combined Virufy-NoCoCoDa datasets is likely due to the inclusion of Virufy data in the merged subset with NoCoCoDa. Similarly, the Cambridge (Symptomatic) testing dataset shows better results when trained with the Cambridge (Asymptomatic) dataset because the cough sounds are from people in the same region. Furthermore, datasets may differ in terms of recording quality, equipment used, and environmental factors, leading to difficulties in transferring learned features to new datasets.

Our findings on poor cross-datasets performance align with previous studies. \citet{akman2022evaluating} note the superior performance of their classifier on trained datasets but acknowledge challenges in classifying datasets differing from the training set. This observation aligns with the concerns raised by \citet{coppock2021covid} regarding pervasive bias in current COVID-19 audio datasets, allowing machine learning models to infer COVID-19 status not just from unique audio biomarkers but also from other dataset correlations, including comorbidity, geographical, ethnic factors, and background noise.

\subsection{Classification performance of proposed method for the combined dataset}

Given the intricate nature of feature characteristics, it was a strategic choice to forgo applying the RFECV technique for the Combined dataset. It is an acknowledgment that there can be significant variability in data, even within the same area. In one dataset, the best features for precise COVID-19 diagnosis may not be the same in another. This focus on flexibility and data-specific optimization highlights a dedication to accurate categorization in the context of heterogeneous datasets. Meanwhile, the cornerstone of our model refining approach has been BO. By providing a data-driven and intelligent method for Hyper-parameter tuning, BO allows the model to adjust its configurations dynamically. This improves prediction accuracy and establishes the model as a reliable and adaptable tool for COVID-19 diagnosis. It is evidence of the incorporation of cutting-edge methods to guarantee optimal performance.

Furthermore, during the training phase, we have integrated SMOTE. By fostering a more representative training dataset, this data augmentation technique lessens the effects of class imbalance and improves the model's capacity to identify patterns in both major and minor classes. The TM technique was also used - it modifies the decision threshold in real-time according to the ROC-AUC scores. With the help of this dynamic adaptation, the model's sensitivity and specificity are precisely balanced, resulting in predictions that are precise and well-informed.

\begin{figure*}[!ht]
     \centering
     \begin{subfigure}[b]{0.49\textwidth}
         \centering
         \includegraphics[width=\textwidth]{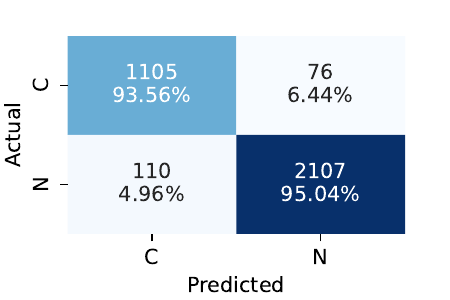}
         \caption{}
         \label{fig:Com1}
     \end{subfigure}
     \hfill
     \begin{subfigure}[b]{0.49\textwidth}
         \centering
         \includegraphics[width=\textwidth]{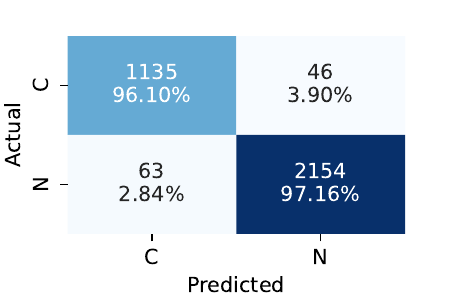}
         \caption{}
         \label{fig:Com2}
     \end{subfigure}
     \caption{The confusion matrices for (a) the 
DNDT classifier and (b) the DNDF classifier, using the combined dataset, are generated through 10-fold cross-validation. In these matrices, ``C" signifies COVID-19 Positive, while ``N" signifies Non COVID-19 cough instances.}
     \label{fig:confusionmatrixscombinedDNDT&DNDF}
\end{figure*}

Table \ref{tab: tab10} presents a comparative analysis that offers a quantitative assessment of our suggested techniques for COVID-19 identification from cough samples, emphasizing the Combined dataset. The DNDF+BO+SMOTE+TM method consistently performs better than the DNDT+BO+SMOTE+TM method across all evaluation metrics. Specifically, the DNDF+BO+SMOTE+TM method exhibits higher values for Accuracy (0.97), ROC-AUC score (0.97), Precision (0.95), Recall (0.96), F1 score (0.96), and Specificity (0.97). Figure \ref{fig:confusionmatrixscombinedDNDT&DNDF} depicts the confusion matrices for the combined dataset, showcasing the effectiveness of our proposed methodologies for accurate COVID-19 classification from cough sounds using the DNDT and DNDF classifiers. The matrices for the DNDT and DNDF classifiers, obtained through 10-fold cross-validation, are presented in Figures~\ref{fig:Com1} and \ref{fig:Com2}, respectively. The DNDT classifier accurately identifies 1,105 out of 1,181 COVID-19 positive cough samples and 2,107 out of 2,217 cough samples from healthy individuals. Similarly, the DNDF classifier predicts 1,135 out of 1,181 COVID-19 positive cough samples and 2,154 out of 2,217 cough samples from healthy individuals.

\begin{table}[!ht]
\centering
\caption{A comparison between our proposed methods for the combined dataset.}
\label{tab: tab10}
\begin{tabular}{llcccccc}
\hline
Dataset                   & Method        & Accuracy           & AUC          & Precision       & Recall        & F1-Score & Specificity          \\ \hline\hline
\multirow{2}{*}{Combined} & DNDT+BO+SMOTE+TM & 0.95         & 0.94         & 0.91          & 0.94 & 0.92                                               & 0.95          \\ & DNDF+BO+SMOTE+TM & \textbf{0.97} & \textbf{0.97} & \textbf{0.95} & \textbf{0.96} & \textbf{0.96}                                      & \textbf{0.97} \\ \hline

\multicolumn{8}{l}{- \textbf{Bold} values represent the highest scores.}

\end{tabular}
\end{table}

\section{Conclusion}
\label{sec:conclusion}

In conclusion, this paper presents a novel method using deep neural decision trees and forests to classify COVID-19 based on cough sound analysis. Our method entails feature extraction, feature selection through RFECV, Hyper-parameter optimization via Bayesian Optimization, SMOTE for data augmentation during training, and establishing an optimal threshold using threshold moving. Extensive performance evaluation is conducted across five diverse datasets- Cambridge, Coswara, COUGHVID, Virufy, and the combined Virufy with NoCoCoDa datasets. We present two ML-based methods for COVID-19 classification using cough sound data: DNDT+RFECV+BO+SMOTE+TM and DNDF+RFECV+BO+SMOTE+TM. On all datasets, our strategies consistently outperform the state-of-the-art techniques. They obtained noteworthy AUC scores of 0.97, 0.98, 0.92, 0.93, 0.99, and 0.99 for the Cambridge Asymptomatic, Cambridge Symptomatic, Coswara, COUGHVID, Virufy, and Virufy merged with NoCoCoDa datasets, respectively, and precision scores of 1, 1, 0.72, 0.93, 1, and 1, respectively.
These findings highlight the efficiency of our proposed methods for correctly classifying COVID-19 from cough sound data, with potentially useful implications for the early detection and tracking of the illness.


In our comprehensive cross-datasets analysis across five diverse datasets, we observed that classification performance tends to decrease in cross-datasets evaluations. This is primarily due to the variations in recording quality, equipment used, demographic differences, and ethnic and environmental factors, which make it difficult to transfer learned features from one dataset to another. However, by amalgamating these five datasets into a unified dataset, we were able to improve performance significantly. Specifically, our proposed approach, using the DNDF classifier with strategy 5, demonstrated promising results with an Accuracy of 0.97, AUC of 0.97, Precision of 0.95, Recall of 0.96, F1-score of 0.96, and Specificity of 0.97.

This insight highlights the trade-off between specificity and generalizability in machine learning models. While cross-dataset evaluations expose the challenges of transferring learned patterns across datasets with differing characteristics, training on a combined dataset enables the model to generalize better, yielding improved performance across a broader range of data. This underscores the importance of dataset diversity in training machine learning models, particularly in domains like COVID-19 detection from audio signals, where data variability is high.


\section*{Declaration of competing interest}
\label{Declaration of Competing Interest}

All authors declare that there is no conflict of interest in this work.

\section*{Acknowledgments}
\label{sec:Acknowledgement}

We thank the University of Cambridge team for supplying the Cambridge dataset and Madison Cohen-McFarlane of Carleton University for sharing the NoCoCoDa dataset. Please note that the University of Cambridge and Carleton University bear no responsibility for the outcomes and findings presented in this article.

\bibliographystyle{elsarticle-num-names} 
\bibliography{reference}

\end{document}